\documentclass[preprint]{aastex}

\received{}
\accepted{}

\slugcomment{8 May 2002 revision}

\author{
Gordon T. Richards\altaffilmark{1},
Daniel E. Vanden Berk\altaffilmark{2},
Timothy A. Reichard\altaffilmark{1},
Patrick B. Hall\altaffilmark{3,4},
Donald P. Schneider\altaffilmark{1},
Mark SubbaRao\altaffilmark{5},
Anirudda R. Thakar\altaffilmark{6}, and
Donald G. York\altaffilmark{5,7}
}

\altaffiltext{1}{Department of Astronomy and Astrophysics, The Pennsylvania State University, University Park, PA 16802}
\altaffiltext{2}{Fermi National Accelerator Laboratory, P.O. Box 500, Batavia, IL 60510}
\altaffiltext{3}{Princeton University Observatory, Princeton, NJ 08544}
\altaffiltext{4}{Pontificia Universidad Cat\'{o}lica de Chile, Departamento de Astronom\'{\i}a y Astrof\'{\i}sica, Facultad de F\'{\i}sica, Casilla 306, Santiago 22, Chile}
\altaffiltext{5}{The University of Chicago, Department of Astronomy and Astrophysics, 5640 S. Ellis Ave., Chicago, IL 60637}
\altaffiltext{6}{Department of Physics and Astronomy, The Johns Hopkins University, 3701 San Martin Drive, Baltimore, MD 21218}
\altaffiltext{7}{The University of Chicago, Enrico Fermi Institute, 5640 S. Ellis Ave., Chicago, IL 60637}

\begin{document}

\title{Broad Emission Line Shifts in Quasars: An Orientation Measure
for Radio-Quiet Quasars?}

\begin{abstract}

Using a sample of 3814 quasars from the Early Data Release of the
Sloan Digital Sky Survey, we confirm that high-ionization, broad,
emission lines such as \ion{C}{4} are significantly blueshifted with
respect to low-ionization, broad, emission lines, such as \ion{Mg}{2},
which are thought to be close to the systemic redshift.  We examine
the velocity shifts of the \ion{Mg}{2} and \ion{C}{4} emission lines
with respect to [\ion{O}{3}] and \ion{Mg}{2}, respectively.
\ion{C}{4} emission line peaks have a range of shifts from a redshift
of $500\,{\rm km\,s^{-1}}$ to blueshifts well in excess of $2000\,{\rm
km\,s^{-1}}$ as compared to \ion{Mg}{2}.  We confirm previous results
that suggest an anti-correlation between the shift of the \ion{C}{4}
emission line peak and the rest equivalent width of the \ion{C}{4}
emission line.  Furthermore, by creating composite quasar spectra as a
function of \ion{C}{4} shift, we are able to study, in detail, the
profiles of the line as a function of velocity shift.  We find that
the apparent shift of the \ion{C}{4} emission line peak is not a shift
so much as it is a lack of flux in the red wing for the composite with
the largest apparent shift.  This observation should strongly
constrain models for the broad emission line region in quasars.  The
emission line blueshift and equivalent width of \ion{C}{4} are also
discussed in light of the well-known anti-correlation between the
equivalent width of \ion{C}{4} emission and continuum luminosity,
otherwise known as the Baldwin Effect.  We further discuss the
\ion{C}{4} emission line shift as a function of other quasar
properties such as spectral index, radio and X-ray detection.  We find
a possible correlation between the \ion{C}{4} emission line shifts and
the radio properties of the quasars that is suggestive of orientation
as the cause of the \ion{C}{4} velocity shifts.  Finally, we explore
whether the \ion{C}{4} emission line blueshifts correlate with the
presence of broad absorption line absorption troughs or with narrow,
``associated'' absorption and how these might be related to
orientation.

\end{abstract}


\section{Introduction}

It has long been known that the redshifts derived from different
quasar emission lines often do not agree with each other within
typical measurement errors.  \markcite{gas82}{Gaskell} (1982) performed the first
detailed study of this phenomenon and concluded that high-ionization,
broad, emission lines such as \ion{C}{4} are shifted by a few hundred
${\rm km\,s^{-1}}$ blueward of the redshifts determined from
low-ionization, broad, emission lines such as \ion{Mg}{2}.  Subsequent
studies \markcite{wil84,ecb+89,cor90,tf92,mrr+99,smd00}({Wilkes} 1984; {Espey} {et~al.} 1989; {Corbin} 1990; {Tytler} \& {Fan} 1992; {McIntosh} {et~al.} 1999; {Sulentic}, {Marziani}, \&  {Dultzin-Hacyan} 2000) have confirmed
these shifts beyond any shadow of a doubt.  Using a sample of over
2200 active galactic nuclei (AGN), \markcite{vrb+01}{Vanden Berk} {et~al.} (2001) demonstrated that
these effects are also present in the sample of quasars from the Sloan
Digital Sky Survey (SDSS;\markcite{yor+00}{York} {et~al.} 2000).

These shifts affect many of the most important issues in
quasar-related science (both directly and indirectly).  The most
obvious area of research that is affected is in the modeling of the
phenomena of AGN, particularly the broad emission line region (BELR;
\markcite{pet97,kro99}{Peterson} 1997; {Krolik} 1999).  BELR models based upon accretion disk winds
\markcite{mc97,psk00}({Murray} \& {Chiang} 1997; {Proga}, {Stone}, \& {Kallman} 2000) will have different dynamics than cloud-based
models \markcite{bm75,cfb80}({Blumenthal} \& {Mathews} 1975; {Capriotti}, {Foltz}, \& {Byard} 1980).  These differences affect the line profiles
and their locations in ways that should be testable.

In addition to investigations of the BELR, investigations regarding
associated absorption \markcite{fwp+86}({Foltz} {et~al.} 1986), broad absorption lines
\markcite{wmf+91}({Weymann} {et~al.} 1991), and the cosmic UV and X-ray background are affected by
the apparent blueshift of the \ion{C}{4} emission line.  We will
comment on each of these issues in turn.

Determining the nature of the so-called ``associated absorber''
\markcite{fwp+86}({Foltz} {et~al.} 1986) population ($z_{\rm abs} \approx z_{\rm em}$), requires
understanding the velocity shift of the \ion{C}{4} emission line in
quasars.  Some associated absorbers are observed to have redshifts
greater than the quasar redshift as determined from the \ion{C}{4}
emission line peak.  However, as many authors have suggested, these
apparent redshifts could simply be the result of the blueshift of the
\ion{C}{4} emission line relative to the systemic redshift of the
quasar.  Given the average observed blueshift of the \ion{C}{4}
emission line, it is clear that the population of truly $z_{\rm abs}
\ge z_{\rm em}$ systems is smaller than it would appear, but it
remains unclear if these systems can be completely explained as
outflows that only appear to be infalling because of the observed
shift of the \ion{C}{4} emission line peak.

In addition to the narrower associated absorbers, emission line shifts
are important for broad absorption line (BAL) quasars.  Previous
investigations of the velocity shifts of quasar emission lines, such
as that of \markcite{cor90}{Corbin} (1990), suggest a correlation between the velocity
shift of the emission lines and the broad absorption line troughs
observed in at least 10\% of all quasars \markcite{wey97}({Weymann} 1997).  However,
\markcite{wmf+91}{Weymann} {et~al.} (1991) saw no such effect.  Whether or not the velocity shifts
are an orientation effect has strong implications for BAL quasars.  If
we can determine the nature of the emission line shifts, we can likely
settle the issue of how BALs are distributed relative to the plane of
the accretion disk that is thought to feed AGN.  Are BAL flows found
preferentially in the plane of the accretion disk, or are the flows
more spherically symmetric in their distribution?

In addition to the more obvious problems associated with our inability
to determine accurate redshifts for quasars, we must also consider the
effect that the apparent inconsistency of emission line redshifts has
on indirect quasar science.  For example, the blueshift of the
\ion{C}{4} emission line causes an over-estimation of the ionizing
flux in the inter-galactic medium, such as is described by
\markcite{mrr+99}{McIntosh} {et~al.} (1999).

The issue of \ion{C}{4} emission blueshift takes on even more
importance if these shifts are correlated with the line of sight
orientation of quasars, or with an orientation-type parameter such as
the opening angle of a disk-wind.  The most obvious ramification of
such a possibility is the ability to determine orientation angles for
radio-quiet quasars: until now, only a small percentage of quasars
with distinct radio properties could be analyzed in terms of
orientation.  If a measure of orientation for radio-quiet quasars
could be found and is robust (whether it be related to the \ion{C}{4}
emission line shifts or not), it would likely revolutionize the study
of quasars.

The format of the paper is as follows.  In \S~2, we describe the data
used in this analysis.  Throughout this paper we will follow
\markcite{sch+02}{Schneider} {et~al.} (2002) and use a cosmology where ${\rm H_o =
50\,km\,s^{-1}\,Mpc^{-1}}, \Omega = 1, \Lambda = 0$.  We also adopt a
quasar-centered coordinate system in which outflows from the quasar
along our line of sight (which appear blueshifted) have {\em positive}
velocities.  Section 3 presents an analysis of the data in terms of
composite quasar spectra, and the radio and X-ray properties of our
sample.  In \S~4, we discuss the implications of our findings for
modeling of the BELR, the Baldwin Effect, the origin of intrinsic
absorption, and orientation measures.  We also give suggestions for
future work.  Finally, \S~5 presents a summary of our results.

\section{Data}

The quasars were selected for spectroscopic followup from the SDSS
imaging survey which uses a wide-field multi-CCD camera
\markcite{gcr+98}({Gunn} {et~al.} 1998); quasar candidates were identified using a preliminary
version of the SDSS Quasar Target Selection Algorithm \markcite{ric+02}({Richards et al.} 2002).
Objects were selected based on both their broad-band SDSS colors
\markcite{fig+96,sto+02}({Fukugita} {et~al.} 1996; {Stoughton} {et~al.} 2002) and as optical matches to radio-detected quasars
from the VLA ``FIRST'' survey \markcite{bwh95}({Becker}, {White}, \& {Helfand} 1995).  All of the emission line
properties are taken directly from the automated 1-D spectroscopic
pipeline (SPECTRO; Frieman et al., private communication;
\markcite{sto+02}{Stoughton} {et~al.} 2002, \S~4.10) outputs that are stored in the public
database.

We study a sample of 3814 bona-fide quasars ($M_i < -23$, with at
least one line broader than $1000\,{\rm km\,s^{-1}}$) from the Early
Data Release (EDR; \markcite{sto+02}{Stoughton} {et~al.} 2002) quasar sample \markcite{sch+02}({Schneider} {et~al.} 2002).
Though the sample is not technically homogeneous, it is adequate for
the scientific goals of this work.  The differences between the EDR
versions of quasar target selection \markcite{sto+02}({Stoughton} {et~al.} 2002) and the final
version of quasar target selection \markcite{ric+02}({Richards et al.} 2002) are largely for
$z>2.2$, whereas our sample is restricted to $z\le2.2$.  In addition,
we cannot think of any reason why the EDR quasar sample would be
biased against (or towards) quasars with large (or small) emission
line blueshifts.

Two sub-samples of quasars were created, one for the study of the
velocity shifts between \ion{C}{4} and \ion{Mg}{2}, the other for the
study of the velocity shifts between \ion{Mg}{2} and [\ion{O}{3}].
These samples contain 794 and 417 quasars, respectively.  

For the \ion{Mg}{2} -- [\ion{O}{3}] sample we take the redshift
determined from [\ion{O}{3}] to be the systemic (center of mass)
redshift, since previous studies have shown that the redshifts
determined from narrow forbidden lines agree well with redshifts
determined from stellar absorption features \markcite{vo85}({Vrtilek} \& {Carleton} 1985) and
\ion{H}{1} 21\,cm emission \markcite{hgp87}({Hutchings}, {Gower}, \& {Price} 1987) in the host galaxies of AGN;
see the discussions in \markcite{tf92}{Tytler} \& {Fan} (1992) and \markcite{mrr+99}{McIntosh} {et~al.} (1999).  To ensure
that both the \ion{Mg}{2} and [\ion{O}{3}] emission lines are in each
of the spectra considered, the redshifts were restricted to $0.415 \le
z \le 0.827$.

For the \ion{C}{4} -- \ion{Mg}{2} sample, our spectra do not cover
[\ion{O}{3}] and we take the redshift determined from \ion{Mg}{2} to
be the systemic redshift, since low-ionization broad lines such as
\ion{Mg}{2} have been shown to have a smaller velocity offset from the
forbidden narrow lines than do high-ionization broad lines
\markcite{tf92}(e.g., {Tytler} \& {Fan} 1992).  In this sample, the redshifts were restricted
to $1.54 \le z \le 2.2$.  These redshift restrictions ensure that both
the \ion{C}{4} and \ion{Mg}{2} emission lines are observed in each of
the spectra considered (which span the range from $3800\,{\rm \AA}$ to
$9200\,{\rm \AA}$ at a resolution of $\approx1800$).  The sample was
further restricted by removing all broad absorption line quasars,
since the BAL troughs hinder accurate measurement of the line peaks.
BAL quasars were defined according to the computation of the
traditional \markcite{wmf+91}{Weymann} {et~al.} (1991) ``balnicity index'' using the composite
quasar spectrum from \markcite{vrb+01}{Vanden Berk} {et~al.} (2001) to define the continuum level.
This list of BAL quasars is not yet final; some weaker BALs may move
into or out of the sample.  However, all strong BALs (balnicity
$\gtrsim 500\,{\rm km\,s^{-1}}$) should be accounted for.  Since the
classification of weak BALs is subjective anyway, we have proceeded
with our analysis using our current list of BALs.

The ``systemic'' redshifts used in this analysis were determined using
only the [\ion{O}{3}] or \ion{Mg}{2} emission lines, respectively.
Since our analysis is dependent upon how SPECTRO measures the peaks of
these lines, we describe some details of the process.  SPECTRO
measures two sets of line parameters. The first are detected peaks in
the spectrum which best match a list of common emission lines.  Once
an object's redshift has been determined, either from the detected
emission lines or from cross-correlation with template spectra, a
second set of line measurements are made at the expected positions of
a more comprehensive set of spectral features. In both cases a single
Gaussian is fit to the line. The first set are identified in the
database as {\em found} and the second as {\em measured}.

The following concerns the {\em measured} line parameters, which are
used in this paper.  The code determines which lines are to be fit
individually and which are to be fit as blends.  The width, height and
center of the line or blend of lines are then fit by a modified
Levenberg-Marquardt algorithm.  To enhance the robustness of the code,
only a single shift from the expected positions is used for blended
lines.  The shift of the line center from the expected position is
constrained to be $\pm2000\,{\rm km\,s^{-1}}$ by multiplying the
$\chi^2$ values by a steep ramp function beyond this range.  In some
cases the expected positions are not the redshifted laboratory
wavelengths, but the redshifted {\em average} line positions as
determined by the composite SDSS quasar spectrum \markcite{vrb+01}({Vanden Berk} {et~al.} 2001); the
line wavelengths relevant here are \ion{C}{4} $\lambda 1545.86$ and
\ion{Mg}{2} $\lambda 2800.32$ (which are doublets, but quasar emission
line widths are broad enough that the doublet is not resolved).  These
wavelengths are used to compute the redshifts as reported in the SDSS
database.  In this analysis, we will determine the redshift for each
individual line based on the SPECTRO {\em measured} line peaks, but we
will use the laboratory rest-frame wavelengths (see \markcite{vrb+01}{Vanden Berk} {et~al.} 2001)
as the expected line positions.
 
\section{Analysis}

\subsection{Distribution of Velocity Shifts}

\begin{figure}[p]
\epsscale{1.0}
\plotone{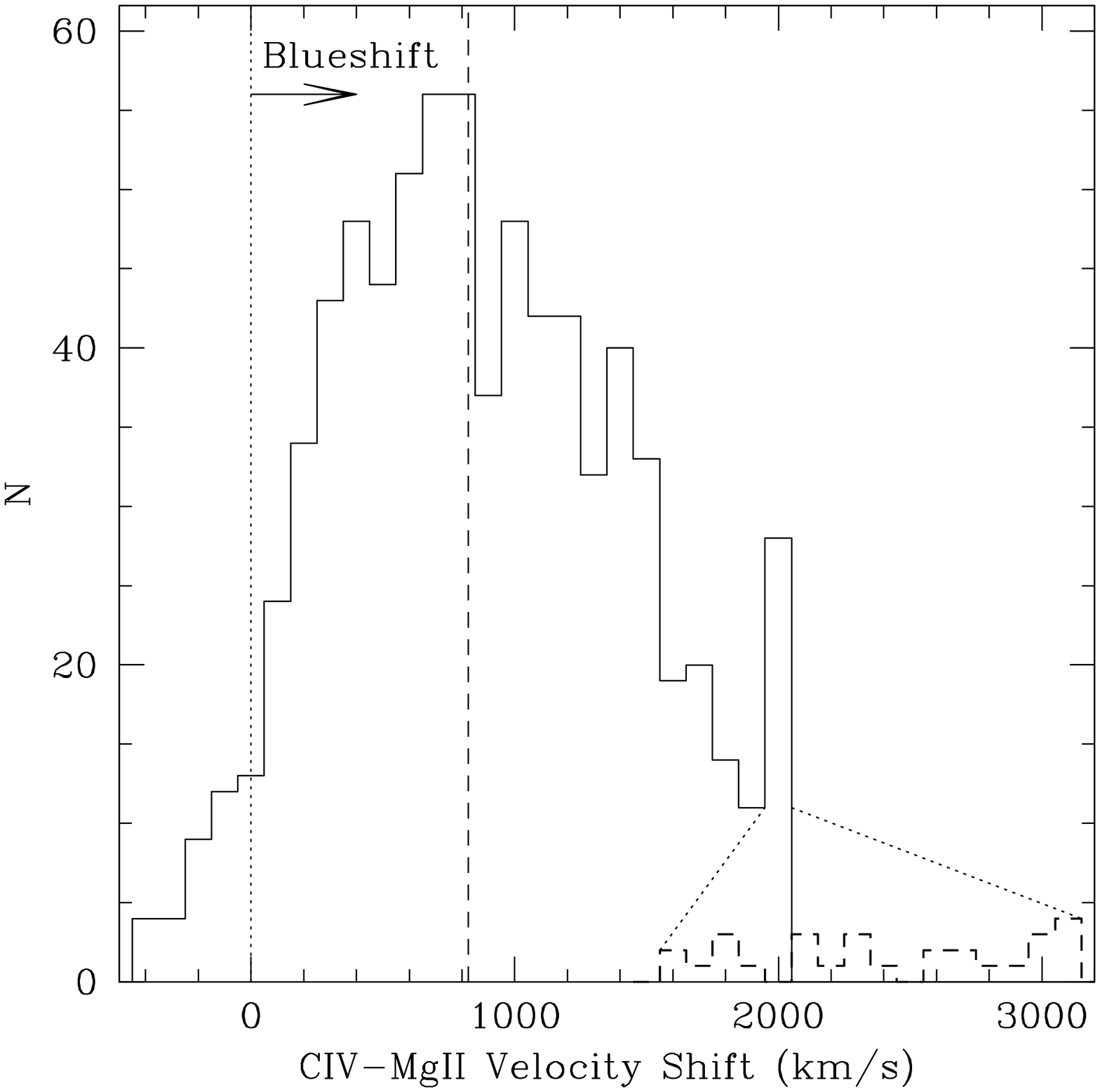}
\caption{Distribution of the 794 \ion{C}{4} emission line velocity
shifts with respect to the \ion{Mg}{2} emission line redshift (solid
line).  The median is $824\,{\rm km\,s^{-1}}$ (vertical, dashed line),
and the dispersion is $\pm511\,{\rm km\,s^{-1}}$.  Positive velocities
indicate a blueshift of \ion{C}{4} with respect to \ion{Mg}{2}.  The
automated code imposes an artificial limit of $2000\,{\rm km s^{-1}}$,
which causes a pile-up at this velocity.  The true range may extend as
high or higher than $3000\,{\rm km s^{-1}}$ when the shifts greater
than $1950\,{\rm km\,s^{-1}}$ are recomputed by hand (28 objects;
dashed histogram).  Errors for individual objects can be
$\gtrsim500\,{\rm km\,s^{-1}}$, but should be much smaller in the
ensemble average.  The largest errors will be found in quasars with
large \ion{C}{4} shift (because the lines are weaker, see
\S~\ref{sec:compspec}) and in quasars with lower than average
signal-to-noise spectra. \label{fig:fig1}}
\end{figure}

The velocity distribution of \ion{C}{4} emission line shifts with
respect to \ion{Mg}{2} is quite large --- as can be seen in
Figure~\ref{fig:fig1}\footnote{Recall that, in this paper, positive
velocities refer to blueshifts with respect to the quasar rest frame,
whereas negative velocities indicate redshifts.}.  We find the shifts
by taking the {\em measured} line peaks for \ion{C}{4} and \ion{Mg}{2}
from the SDSS database and using the laboratory rest-frame wavelengths
of the \ion{C}{4} and \ion{Mg}{2} doublets, ($1549.06$ and
$2798.75\,{\rm \AA}$, respectively) to determine the redshift of each
line.  The median of the distribution is found to be $824\,{\rm
km\,s^{-1}}$ with a dispersion of $\pm511\,{\rm km\,s^{-1}}$.  For
individual quasars, the error in the \ion{C}{4} blueshift with respect
to \ion{Mg}{2} can be $\sim500\,{\rm km\,s^{-1}}$ ($2\,\sigma$).  Spot
checking of good quality spectra to determine the blueshift by hand
supports this conclusion.  We do not believe that this error is skewed
towards either positive or negative shifts, so the errors in the
ensemble average should be much lower.  In general, the errors will be
biggest for those quasars with the largest \ion{C}{4} blueshifts,
since we find that those quasars have the weakest \ion{C}{4} emission
lines; see \S~\ref{sec:errors} for a more detailed discussion
regarding errors.

\begin{figure}[p]
\epsscale{1.0}
\plotone{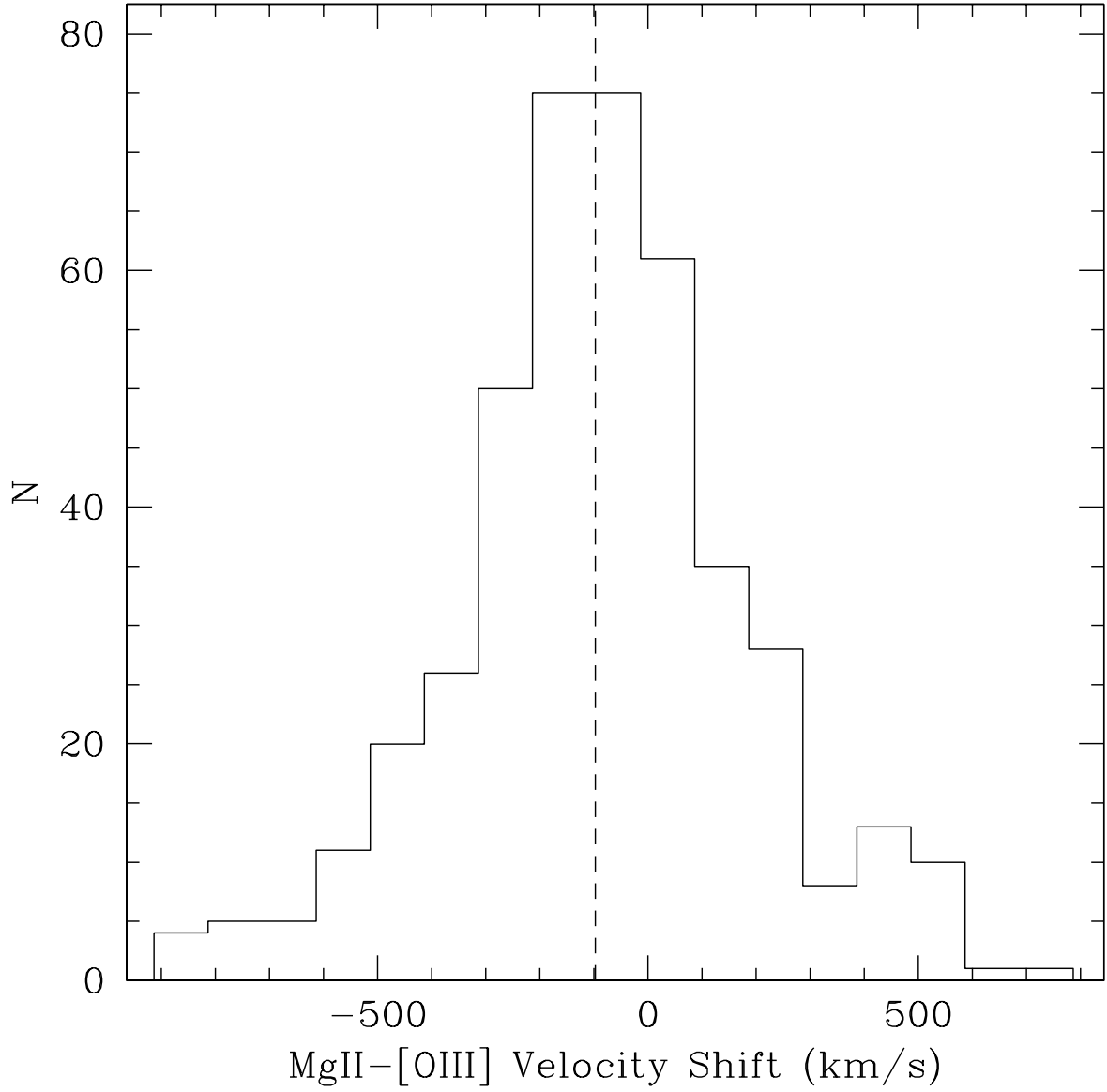}
\caption{Distribution of 417 \ion{Mg}{2} emission line peak velocity
shifts with respect to [\ion{O}{3}].  The median is $-97\,{\rm
km\,s^{-1}}$ (vertical, dashed line), and the dispersion is
$\pm269\,{\rm km\,s^{-1}}$.  Positive velocities indicate a blueshift
of \ion{Mg}{2} with respect to [\ion{O}{3}].\label{fig:fig2}}
\end{figure}

In a similar manner to the \ion{C}{4} -- \ion{Mg}{2} shift,
Figure~\ref{fig:fig2} shows the distribution of velocity differences
between the \ion{Mg}{2} and [\ion{O}{3}] emission line peaks.  Again,
we use the {\em measured} line peaks from the SDSS database, and the
laboratory wavelengths of \ion{Mg}{2} doublet and the reddest of the
[\ion{O}{3}] triplet ($2798.75$ and $5008.24\,{\rm \AA}$,
respectively) in order to determine the redshift of each emission line
and the blueshift between the emission lines.  The peak of this
distribution is near zero, but has an overall median shift of
$-97\,{\rm km\,s^{-1}}$ with a spread of $269\,{\rm km\,s^{-1}}$.
This shift is about half as large as that found by \markcite{vrb+01}{Vanden Berk} {et~al.} (2001), who
found a velocity offset of $-161\,{\rm km\,s^{-1}}$ for
\ion{Mg}{2}\footnote{Note the sign difference between this paper and
\markcite{vrb+01}{Vanden Berk} {et~al.} (2001).  The $+161\,{\rm km\,s^{-1}}$ quoted by \markcite{vrb+01}{Vanden Berk} {et~al.} (2001)
becomes $-161\,{\rm km\,s^{-1}}$ in the notation used herein.};
\markcite{tf92}{Tytler} \& {Fan} (1992) found ambiguous results for the velocity shift of the
\ion{Mg}{2} emission line ($200\,{\rm km\,s^{-1}}$ and $-380\,{\rm
km\,s^{-1}}$ with respect to [\ion{O}{2}] for two different samples,
respectively).  In our sample, the \ion{Mg}{2} emission line, while
quite close to what is thought to be the systemic redshift in the
ensemble average, can have significant deviations in individual
quasars (modulo measurement errors).

Other authors \markcite{mrr+99}(e.g., {McIntosh} {et~al.} 1999) have also studied the shift of
the H$\beta$ emission line.  We are not able to do so using the
automated reductions of the SDSS spectra.  The SPECTRO pipeline treats
H$\beta$ and the nearby [\ion{O}{3}] triplet as a blend with the same
redshift (fortunately, redshifts determined from H$\beta$ are
typically similar to redshifts determined from [\ion{O}{3}]).  An
analysis of the shift of the H$\beta$ emission line with respect to
[\ion{O}{3}] would thus require a substantial modification to the
SPECTRO pipeline and a reanalysis of all of the spectra: a task which
is well beyond the scope of this paper.

\subsection{Composite Spectra\label{sec:compspec}}

\subsubsection{Sub-sample Definition}

To assess the differences between the spectra of quasars with both
large and small \ion{C}{4} -- \ion{Mg}{2} redshift differences, we
created a set of four composite quasar spectra by combining the
spectra of quasars with similar velocity shifts.  We first sorted the
list of quasars according to their \ion{C}{4} -- \ion{Mg}{2} redshift
differences and then divided this list into four roughly equal lists
of $\sim200$ quasars each.  Composite quasar spectra were created by
combining the spectra (using the geometric mean, which is appropriate
for power-law spectra) in each of these four samples following the
procedure of \markcite{vrb+01}{Vanden Berk} {et~al.} (2001).  Table~\ref{tab:tab1} gives the velocity
offset properties of these samples (denoted A, B, C, and D); it
includes the number of objects (the number within the FIRST survey
area is given in parentheses), the average velocity offset of the
\ion{C}{4} emission line in the composite spectrum, along with the
average, minimum and maximum velocity offsets of the \ion{C}{4}
emission line in the individual spectra.  Table~\ref{tab:tab2} gives
other properties of the samples, including the average redshift, the
average absolute magnitude in the $i$-band, the average spectral index
($f_{\nu} \propto \nu^{\alpha}$) of the optical continuum, the number
and fraction of radio and X-ray detected sources, along with the
equivalent width and the full-width at half-maximum of the \ion{C}{4}
emission line (in \AA).

\subsubsection{Velocity Errors\label{sec:errors}}

We have binned the data into four different samples, in part, to
minimize the effect of any errors in the emission line blueshift
determination.  Errors can cause objects move from one bin to the
next, but it is unlikely that many objects from sample A will move
into sample D (and vice versa) as a result of errors.  Also, that the
offset between the peaks of the \ion{Mg}{2} emission lines for the
composite spectra in \S~\ref{sec:specanal} is small suggests that the
errors in the blueshift measurements are not a problem.  Finally, we
have measured the blueshift of \ion{C}{4} by hand (using the mode of
the upper 50\% of the line) in each of the composites.  We find that
the blueshifts are $197, 606, 1003$, and $1526\,{\rm km\,s^{-1}}$, for
composites A, B, C, and D, respectively.  These values are in
excellent agreement with the average blueshifts of the individual
quasars that contribute to each composite as can be seen by comparing
columns 3 and 4 in Table~1, thus we are confident that measurement
errors do not significantly effect our results.

Note that, as discussed above, the measured emission line centers have
an upper limit of $2000\,{\rm km\,s^{-1}}$ for the velocity difference
of the line center from the accepted redshift of the quasar.  Some of
the \ion{C}{4} peaks are shifted by more than this $2000\,{\rm
km\,s^{-1}}$, thus there is a ``pile-up'' at this velocity in the
database.  Future versions of SPECTRO will allow for shifts as large
as $3000\,{\rm km\,s^{-1}}$.  For those quasars affected by this
limitation in SPECTRO (28 in all), we have measured the velocity
shifts by hand; the \ion{C}{4} emission line peak was determined using
the mode of the upper 50\% of the line as was the case in
\markcite{vrb+01}{Vanden Berk} {et~al.} (2001).  The dashed histogram in Figure~\ref{fig:fig1} shows
the hand calculated velocity shifts of these quasars.  We find that
the error on these \ion{C}{4} shifts can be quite large and that the
true upper velocity limit may be as high as or higher than $3000\,{\rm
km\,s^{-1}}$.

\subsubsection{Composite Spectra Analysis\label{sec:specanal}}

\begin{figure}[p]
\epsscale{1.0} 
\plotone{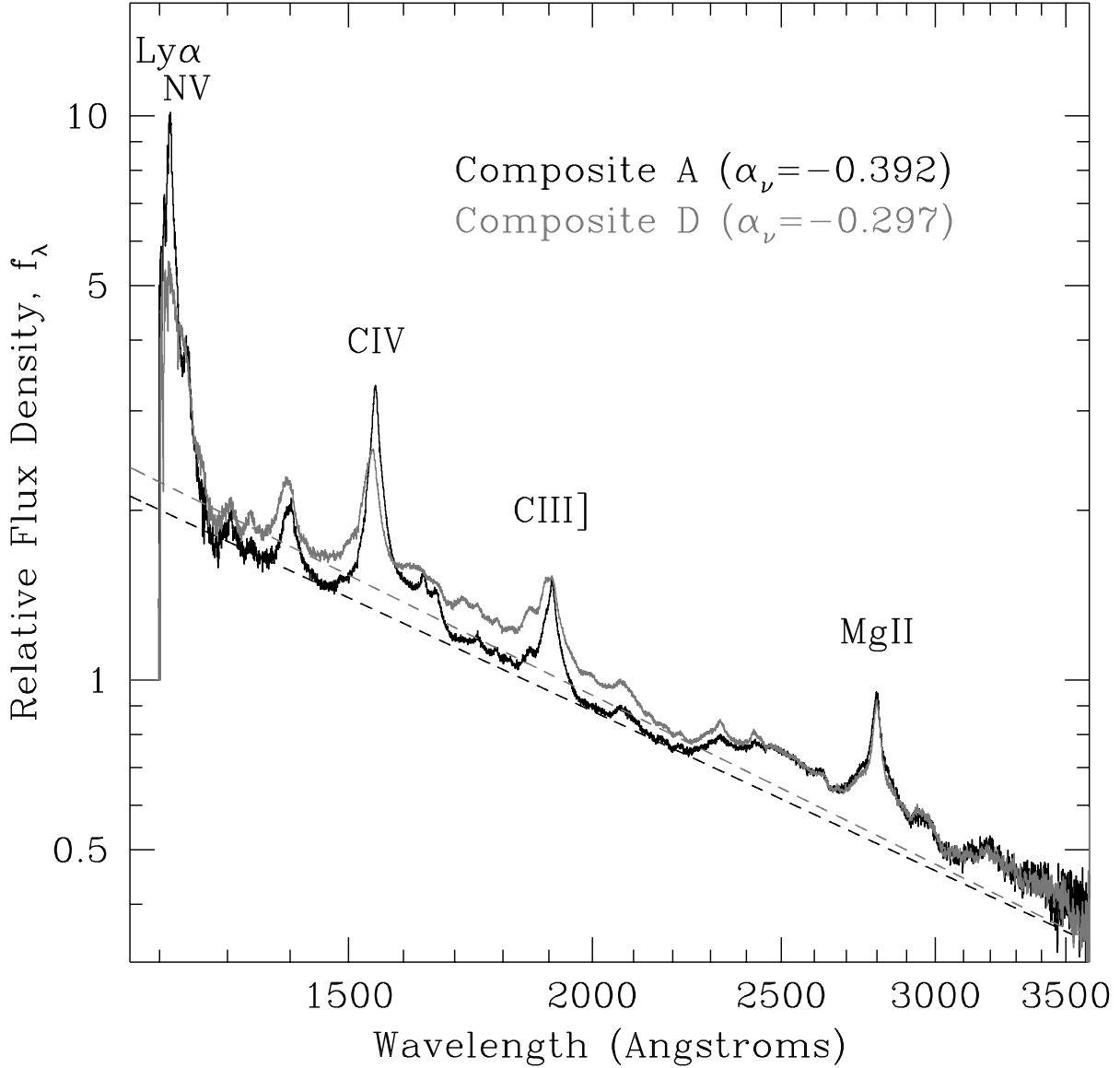}
\caption{Composite spectra of 199 quasars with the smallest \ion{C}{4}
-- \ion{Mg}{2} redshift differences (composite A, black) and 198
quasars with the largest \ion{C}{4} -- \ion{Mg}{2} redshift
differences (composite D, grey).  The spectra have not been normalized
to match each other, see \markcite{vrb+01}{Vanden Berk} {et~al.} (2001) for an explanation of how the
spectra are normalized internally.  The power-law continua have been
computed between $1355\,{\rm \AA}$ and $2200\,{\rm \AA}$ as described
in the text. Major emission lines are indicated; see \markcite{vrb+01}{Vanden Berk} {et~al.} (2001)
for a complete identification of the emission lines. \label{fig:fig3}}
\end{figure}

In Figure~\ref{fig:fig3}, we plot the spectra of the composites whose
inputs had the smallest and largest \ion{C}{4} emission line shifts,
which we hereafter refer to as composites A and D, respectively.  We
also plot power-law continua with $\alpha=-0.392$ and a slightly bluer
$\alpha=-0.297$ for composites A and D, respectively to guide the eye.
The spectral indices were computed using a range of $10\,{\rm \AA}$ at
both $1355\,{\rm \AA}$ and $2200\,{\rm \AA}$.  Neither of these ranges
are ideal continuum windows, see \markcite{vrb+01}{Vanden Berk} {et~al.} (2001), but these are
appropriate values to use in the available wavelength range.  The
composite spectra do not extend far enough into the red to use a
better continuum window.  For comparison, we have also created a
composite quasar spectrum from all of the quasars in the EDR quasar
catalog \markcite{sch+02}({Schneider} {et~al.} 2002), which differs from the sample used by
\markcite{vrb+01}({Vanden Berk} {et~al.} 2001).  The EDR quasar composite has a spectral index of
$\alpha=-0.410$ when measured in the manner described above.  Note,
however, that BALs, which may tend to be redder than average, are not
excluded from the EDR composite quasar spectrum.  Since these continua
are clearly not optimal over the entire spectrum, we will use local
continua for any further analysis.

\begin{figure}[p]
\epsscale{1.0} 
\plotone{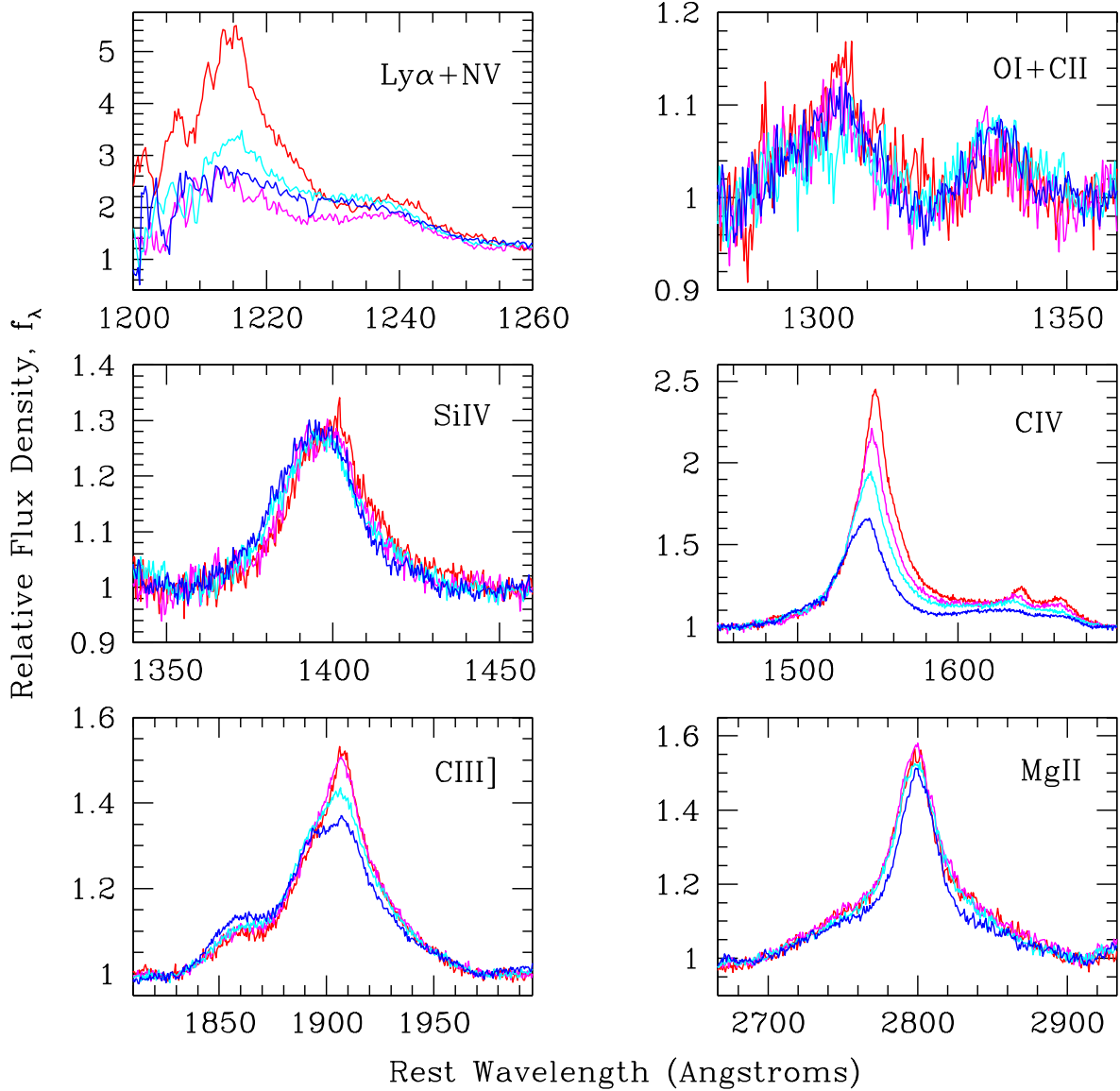}
\caption{The strong emission line regions of the composite spectra,
including the two intermediate \ion{C}{4} shift composite spectra.
Local continuum regions (see text) were used instead of the global
fits from Figure~\ref{fig:fig3}. The spectra, in the order of
increasing \ion{C}{4} shift (composites A, B, C, and D), are plotted
as red, magenta, cyan, and blue, respectively.  Note the striking
difference in the \ion{C}{4} and \ion{C}{3}] profiles.  The labels in
the upper right-hand corner of each panel gives the dominant emission
line(s) in each respective panel; see \markcite{vrb+01}{Vanden Berk} {et~al.} (2001) for a complete
identification of the emission lines.\label{fig:fig4}}
\end{figure}

Figure~\ref{fig:fig4} shows the major emission line regions for all
four of the (normalized) composite spectra (A, B, C, and D) with the
red spectrum having the smallest \ion{C}{4} shift (composite A) and
the blue spectrum showing the largest \ion{C}{4} shift (composite D).
Here we have chosen to normalize the spectra in local continuum
regions.  The continuum regions [($\lambda_{\rm min}:\lambda_{\rm
max}$), in \AA] are (1285:1355), (1285:1355), (1355:1465),
(1465:1695), (1830:1975), (2695:2955) for Lyman-$\alpha$, \ion{O}{1},
\ion{Si}{4}, \ion{C}{4}, \ion{C}{3}], and \ion{Mg}{2}, respectively.

Of particular interest in Figure~\ref{fig:fig4} is the profile of the
\ion{C}{4} emission line as a function of velocity shift.  By simple
visual inspection, we confirm previous results that found an
anti-correlation between the \ion{C}{4} emission line velocity shift
and equivalent width \markcite{cor90}({Corbin} 1990).  Furthermore, after the spectra
have been normalized by the continuum, it is clear that the shift of
the \ion{C}{4} emission line is not so much a shift of the centroid to
the blue as it appears to be a suppression of flux in the red wing.
The blue wing is quite similar in all four composite spectra, whereas
the peak and the red wing are less pronounced as the velocity shift
increases.  We note that this conclusion depends significantly upon
our continuum placement; however, we feel that our procedure is
appropriate; see below for further discussion.

This extinction in the red wing coupled with the similarity of the
blue wings is more complex than is generally reported.  However, such
an effect is not entirely without precedent: see \markcite{smd00}{Sulentic} {et~al.} (2000); also
both \markcite{fk95}{Francis} \& {Koratkar} (1995) and \markcite{cb96}{Corbin} \& {Boroson} (1996) found enhanced Baldwin Effects
(\markcite{bal77}{Baldwin} 1977, also see \S~\ref{sec:baldwin}) in the red wing of
\ion{C}{4}.  That the blueshift is apparently not a blueshift but
rather a deficit of flux in the red wing {\em significantly} changes
the interpretation of the \ion{C}{4} emission line shifts.  The shift
is clearly not the result of an actual change in the (apparent) bulk
velocity flow (i.e., it is not a blueward shift of the entire line
profile); instead, the shift may be caused entirely by obscuration or
suppression of the lowest (or most negative) velocity gas (in the
quasar rest frame), since the blue wings of both the small and large
blueshift composites are co-located.

By dividing rather than subtracting the continuum, we implicitly
assume that the flux in the emission lines is correlated with the
continuum luminosity of the quasars, as expected since the BELR is
powered primarily by photoionization induced by the UV continuum.
This correlation is not perfect; higher-luminosity quasars have lower
equivalent width emission lines (the Baldwin Effect, \markcite{bal77}{Baldwin} 1977).
However, the continuum luminosity of quasars has a range of values at
least an order of magnitude larger than the range of emission line
equivalent widths; for example, see \markcite{bwg89}{Baldwin}, {Wampler}, \& {Gaskell} (1989), Figure 5 and
\markcite{kro99}{Krolik} (1999), Figure 10.2.  If the continuum was completely decoupled
from the lines, the range of values spanned by emission line EQWs
would be the larger of the two, since it would include intrinsic
scatter plus uncorrelated variations produced by different continuum
levels.  As the opposite is true, dividing by the continuum is more
appropriate than subtracting.

\begin{figure}[p]
\epsscale{1.0} 
\plotone{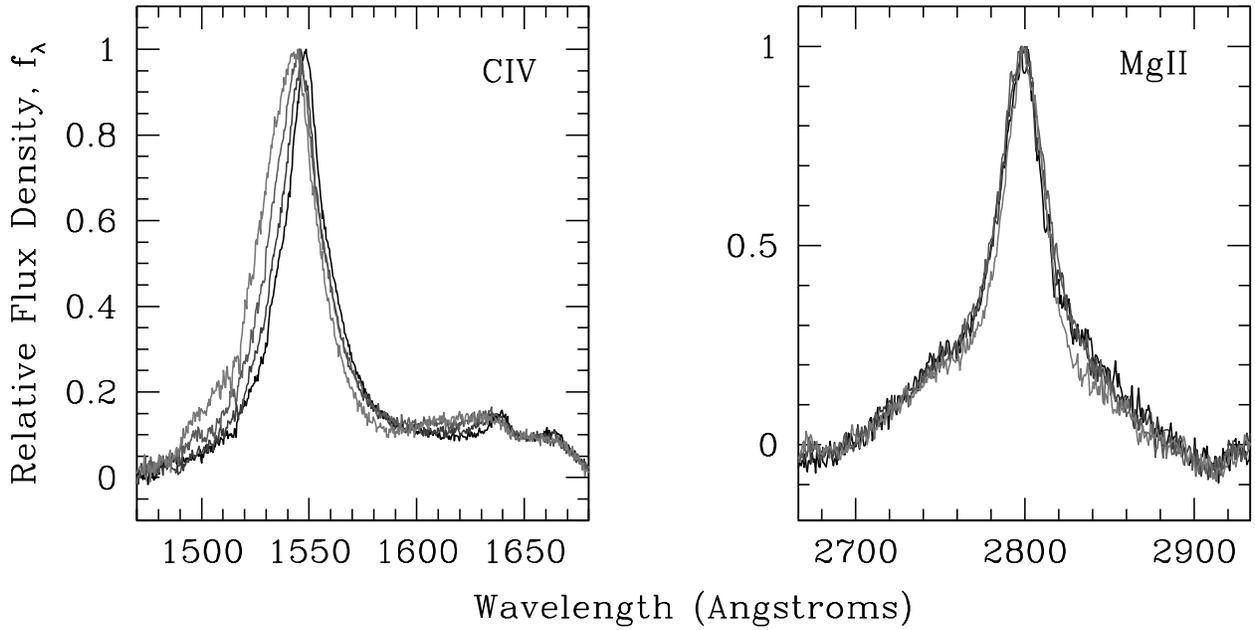}
\caption{The \ion{C}{4} and \ion{Mg}{2} emission line regions
normalized by subtracting instead of dividing the best-fit, local,
power-law continua.  The peaks of the lines have also been scaled to
unity so that the profiles can be compared on the same scale.  Note
that this scaling distorts the relative fluxes between the composites
(see Figure~\ref{fig:fig4}); composite D has more blue flux than red
as compared to composite A, but composite D has less overall blue
flux.  Composite A is shown in black; composite D is in the lightest
grey. \label{fig:fig5}}
\end{figure}

\begin{figure}[p]
\epsscale{1.0} 
\plotone{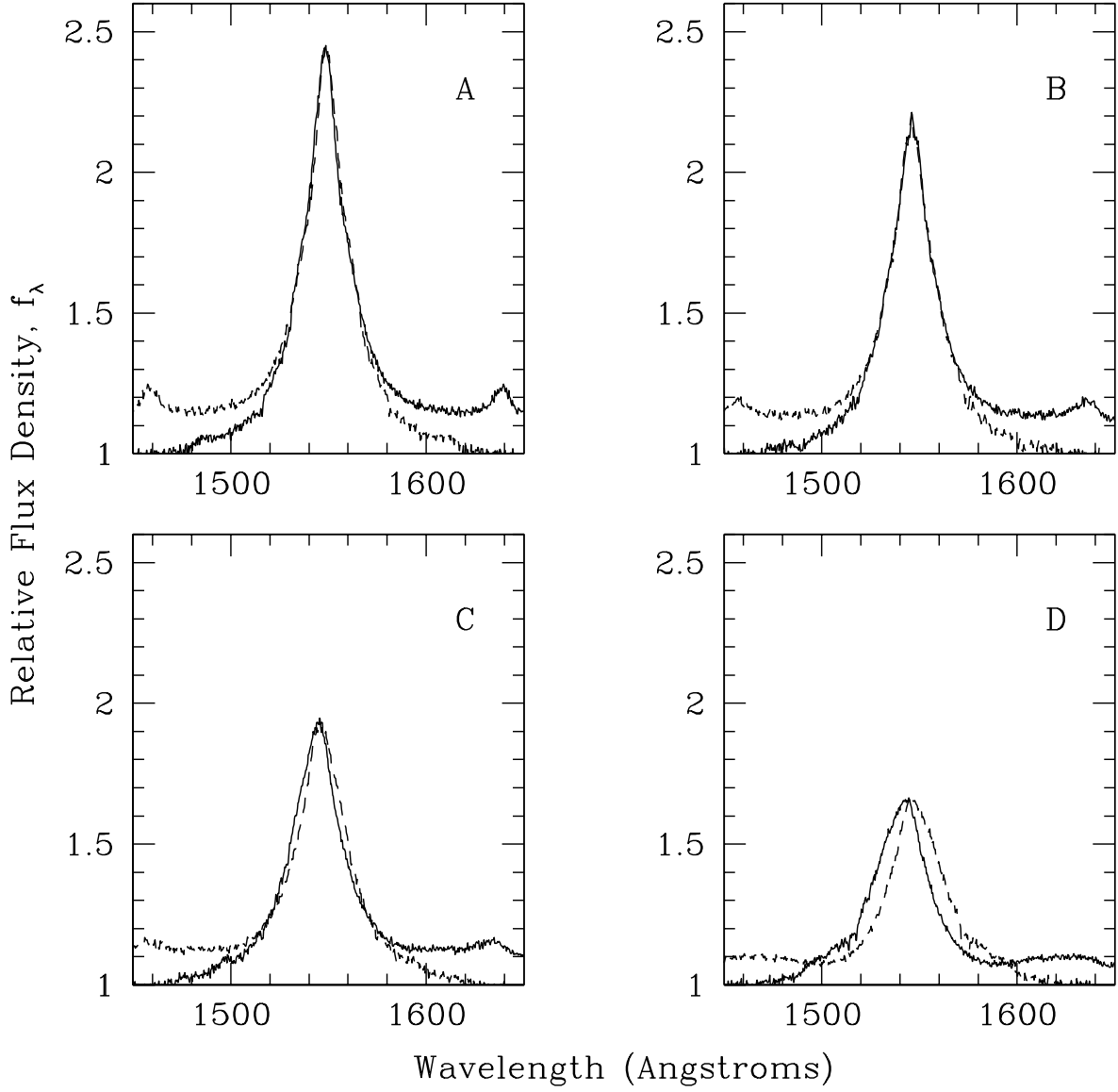}
\caption{Asymmetry of the \ion{C}{4} profiles from composites A, B, C,
and D.  Each \ion{C}{4} profile is inverted around its peak value and
is overplotted as a dashed line.  Significant differences between the
solid and dashed lines indicates significant asymmetry.  Note that we
invert around the observed peak of the line and not the expected peak,
which would emphasize the asymmetries even more as a result of the
apparent blueshifts of the lines. \label{fig:fig6}}
\end{figure}

However, it could be argued that for some cases, such as when trying
to understand the line profiles, it would be more appropriate to
subtract the continuum.  Thus, in Figure~\ref{fig:fig5}, we show the
profiles of \ion{C}{4} and \ion{Mg}{2} after subtracting the continuum
and scaling the emission lines to have the same peak strength.
Figure~\ref{fig:fig5} demonstrates that the \ion{C}{4} line profile is
relatively symmetric in composite A, but is quite asymmetric in
composite D, which has more blue flux than red flux as compared to
composite A (see also Figure~\ref{fig:fig6}).  Note, however, that
the scaling used distorts the relative fluxes between the composites;
although composite D has more blue flux than red as compared to
composite A, composite D has less overall blue flux (see
Figure~\ref{fig:fig4}).  In any case, we emphasize that the
interpretation of the \ion{C}{4} profiles does depend significantly on
how one handles the continuum.

By visual inspection, we find that our composites are similar to those
created by \markcite{wbf+93}{Wills} {et~al.} (1993) except that here we divide the sample by
their \ion{C}{4} -- \ion{Mg}{2} redshift differences instead of by the
FWHM of the \ion{C}{4} emission line.  In addition, we determine the
redshift solely from the \ion{Mg}{2} emission line.  It is therefore
interesting that we see many of the same properties as did
\markcite{wbf+93}{Wills} {et~al.} (1993), such as the anti-correlation between the rest
equivalent width (EQW) and full-width at half maximum (FWHM) of the
\ion{C}{4} emission lines.  Although the rest equivalent widths of the
\ion{C}{4} emission lines are anti-correlated with the FWHM of the
\ion{C}{4} emission lines in our samples (see Table~\ref{tab:tab2}),
it is not clear that composite D, in fact, has the broadest lines.  If
the \ion{C}{4} emission line profile is indeed absorbed in composite D
as compared to composite A, then the FWHM of the emission line is
largely unrelated to the velocity distribution of the emission line
gas.

We further compare to \markcite{wbf+93}{Wills} {et~al.} (1993) by plotting, in
Figure~\ref{fig:fig6}, the \ion{C}{4} profiles of each of our
composites on top of reflections about the peaks of each of the
\ion{C}{4} profiles.  In these plots we have normalized the continua
using the same windows as for \ion{C}{4} above; we find no qualitative
difference in the inverted profiles if instead we use (1480:1620)
Angstroms as the continuum window.  The profiles are inverted around
the observed peaks of the lines and not the expected line centers for
\ion{C}{4}.  If we used the laboratory wavelength as the reflection
point, the asymmetries in the more blueshifted composites would be
even more dramatic.  Whereas \markcite{wbf+93}{Wills} {et~al.} (1993) find that their
intermediate spectra are the most asymmetric, we find that the
asymmetry in our spectra increases with increasing blueshift of the
line peak.

\begin{figure}[p]
\epsscale{1.0} 
\plotone{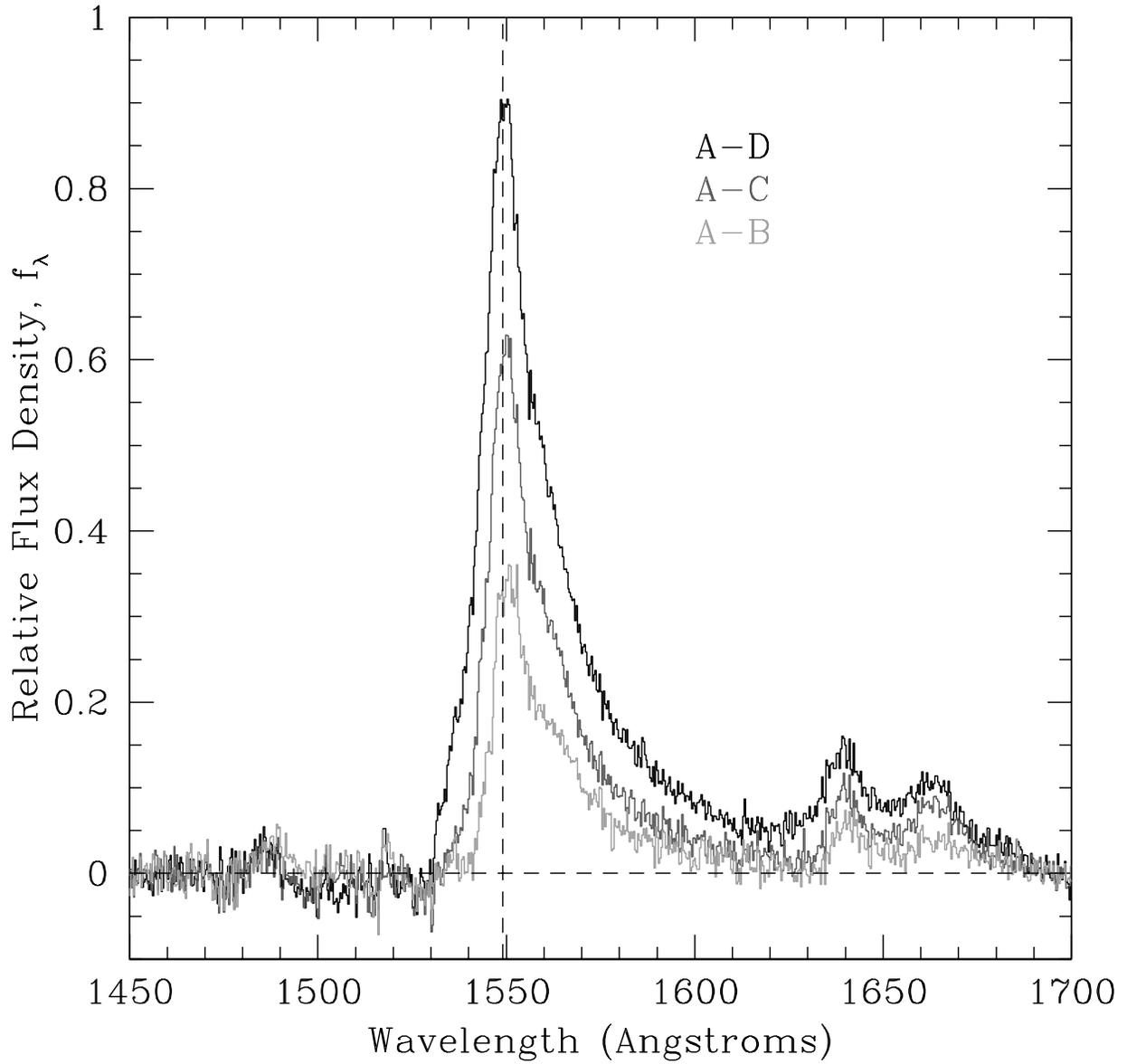}
\caption{Difference spectra in the vicinity of \ion{C}{4}.  The
difference between composites A and D are shown in black, A minus C,
and A minus B are shown in dark and light grey, respectively.  The
vertical dashed line is drawn at $1549.06\,{\rm
\AA}$. \label{fig:fig7}}.
\end{figure}

It is also possible to create difference spectra in the vicinity of
\ion{C}{4}, similar to those made by \markcite{bwf+94}{Brotherton} {et~al.} (1994); these are shown
in Figure~\ref{fig:fig7}.  These spectra have been normalized as
above; the difference between composite A and D is shown in black, A
minus C in dark grey, and A minus B in light grey.  Note that the red
flux missing in composites B, C, and D extends to at least $1600\,{\rm
\AA}$ --- nearly $10,000\,{\rm km\,s^{-1}}$ from the line center.
These \ion{C}{4} difference profiles are considerably more asymmetric
than those of \markcite{bwf+94}{Brotherton} {et~al.} (1994).  This difference is probably the result
of our use of the \ion{Mg}{2} redshift to define the systemic
redshift.  If instead we aligned the peaks of the \ion{C}{4} emission
line in the composite spectra before we created the difference
spectra, we would get more symmetric difference spectra.  If the two
component BELR model discussed by \markcite{wbf+93}{Wills} {et~al.} (1993) is valid, then these
\ion{C}{4} difference spectra and the asymmetries of the line profiles
place significant constraints upon the model; given the shape of the
difference spectra, we would prefer an obscured or suppressed
disk-wind outflow as the preferred explanation for the BELR (see
\S~\ref{sec:discussion}).

Emission lines other than \ion{C}{4} are equally worthy of
consideration.  In particular, we note that, in Figure~\ref{fig:fig4},
the \ion{C}{3}] profile seems to be shifted significantly blueward in
composite D.  This observation is consistent with the findings of
\markcite{ecb+89}{Espey} {et~al.} (1989) and \markcite{cor90}{Corbin} (1990).  However, instead of interpreting
this as a real shift of \ion{C}{3}], we interpret this as an apparent
shift due to an increase in the relative strength of \ion{Si}{3}]
$\lambda 1892$ (see \markcite{vrb+01}{Vanden Berk} {et~al.} 2001) versus \ion{C}{3}], or simply
perhaps as a weakening of \ion{C}{3}].  Alternatively, the changes
could be due to \ion{Fe}{3} (UV34); see \markcite{vw01}{Vestergaard} \& {Wilkes} (2001), \S~5.2.  If the
interpretation that the change in line profile results from a change
in the relative strength of the lines is correct, the velocity
difference for the \ion{C}{3}] peaks between composites A and D
appears to be zero within the measurement errors (however, the errors
are quite large given the severity of the blending).  In any case, it
may still be possible to make use of this emission feature, since, for
the purpose of measuring the shifts, it matters less why the line
shifts than the fact that it shifts in the first place.

The Lyman-$\alpha$+\ion{N}{5} blend in the upper left-hand panel of
Figure~\ref{fig:fig4} is also of interest.  The most obvious result is
that the Lyman-$\alpha$ line is {\em much stronger} for composite A
than for the composite D.  However, this observation is mitigated by
the fact that the line in composite B is also quite weak.  This
discrepancy may be the result of the fact that very few quasars
($\sim10$ per composite) contribute to the composites in this region
of the spectrum, thus the composite Lyman-$\alpha$ profiles (and to a
lesser extent, \ion{N}{5}) are on a much weaker footing than the other
profiles discussed here.

Although more data is needed in the Lyman-$\alpha$+\ion{N}{5} region
of the spectra, we note that a shift of the \ion{N}{5} emission line
is evident between composites A and D (see Figure~\ref{fig:fig4}).  In
composite A, the \ion{N}{5} peak is separated from the red wing of
Lyman-$\alpha$.  However, in composite D, the \ion{N}{5} emission line
blends in with the red wing of Lyman-$\alpha$ --- consistent with a
blueshift of \ion{N}{5}.  The blending is sufficiently strong that we
do not attempt to measure the velocity shift of \ion{N}{5}; however,
visual inspection of Figure~\ref{fig:fig4} indicates that the
\ion{N}{5} shift may be significantly larger than found by
\markcite{vrb+01}{Vanden Berk} {et~al.} (2001) --- a finding which would bring \ion{N}{5} more into
line with the observed correlation between ionization potential and
velocity shift \markcite{vrb+01}({Vanden Berk} {et~al.} 2001).

Other lines, in addition to \ion{C}{4}, show significant changes in
strength in Figure~\ref{fig:fig4}.  Composite D has additional flux in
the \ion{Al}{3} $\lambda 1857$ line in addition to the apparent
increase in \ion{Si}{3}] flux (relative to \ion{C}{3}]), as discussed
above.  Even more striking is the almost complete lack of \ion{He}{2}
$\lambda 1640$ in composite D, meaning that the \ion{He}{2} emission
line region is obscured or suppressed.  Alternatively, the \ion{He}{2}
flux in Composite D may simply be broader, weaker and more blueshifted
than in Composite A.  However, the ``more blueshifted'' part of this
interpretation for \ion{He}{2} is inconsistent with the \ion{C}{4}
results, since we have argued that the \ion{C}{4} blueshift is not a
bulk shift of the line profile to the blue, but rather a lack of red
flux.  Since \ion{He}{2} is a high-ionization line, its absence in the
composite with the largest shift has interesting consequences.  The
blend of \ion{O}{3}], \ion{Al}2{} and \ion{Fe}{2} just redward of this
line also seems to be missing.  Finally, the extended \ion{Fe}{2}
emission (i.e., the $3000\,{\rm \AA}$ bump) appears to be similar in
both composite A and D.

Perhaps as interesting as the differences in the composites in
Figure~\ref{fig:fig4} is one notable lack of difference: \ion{Si}{4}.
Although \ion{Si}{4} is a high-ionization line and is similar in
chemical structure to \ion{C}{4}, we see little difference in the
\ion{Si}{4} profiles between the composites.  There is some evidence
for a shift in the line peaks, but no evidence for dearth of flux in
the red wing.  We have no explanation for this fact, other than to
note that this finding is consistent with previous results for the
Baldwin Effect (see \S~\ref{sec:baldwin}), and that the ionization
potential for creation of \ion{C}{4} is about equal to the ionization
potential for destruction of \ion{Si}{4}.  Thus \ion{Si}{4} emission
will tend to come from larger radii than \ion{C}{4} emission, which
may be a clue to the different behaviors of their profiles.

\begin{figure}[p]
\epsscale{1.0} 
\plotone{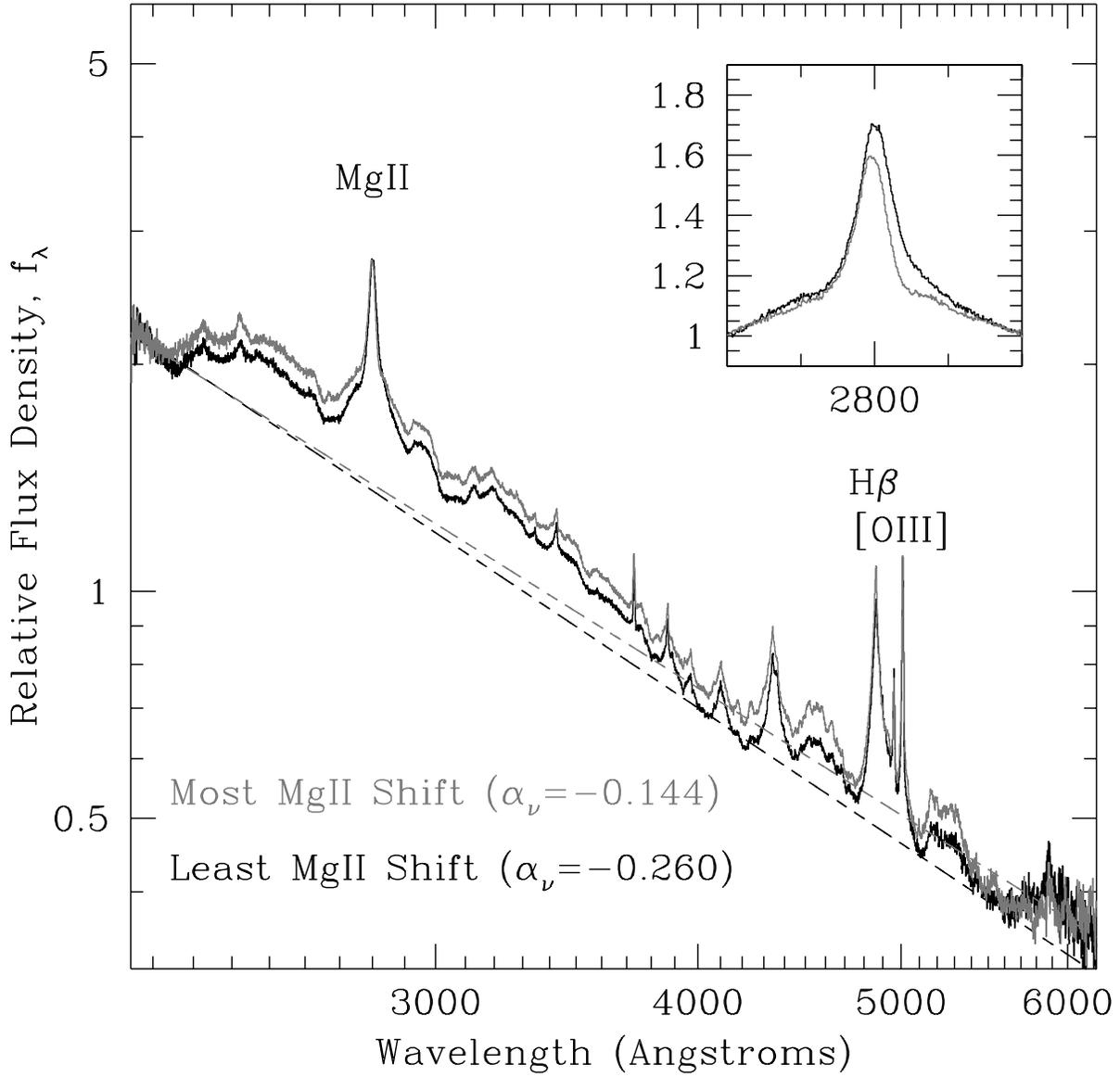}
\caption{Composite spectra of quasars with large (grey) and small or
negative (black) \ion{Mg}{2} -- [\ion{O}{3}] velocity shifts.  Inset
shows \ion{Mg}{2} region on a linear scale after normalizing the
spectra using (local) power-law continuum fits.
\label{fig:fig8}}
\end{figure}

Since redshifts measured from \ion{Mg}{2} are generally thought to be
farther from systemic compared to those measured from [\ion{O}{3}], we
have also created composite spectra covering these two emission peaks
for the sake of comparison.  Figure~\ref{fig:fig2} showed that the
velocity shift of \ion{Mg}{2} as compared to [\ion{O}{3}] ranges from
less than $-500\,{\rm km\,s^{-1}}$ to $500\,{\rm km\,s^{-1}}$.  In
Figure~\ref{fig:fig8}, we plot two composites of the spectra with the
smallest and most negative shifts of \ion{Mg}{2} with respect to
[\ion{O}{3}], in black, and with the largest shifts, in grey.  The
inset of Figure~\ref{fig:fig8} shows the \ion{Mg}{2} emission lines
after the spectra have been normalized by a locally fit power-law
continuum.  As with the \ion{C}{4} profiles, the \ion{Mg}{2} profiles
are in good agreement in the blue wing with an apparent lack of flux
in the red wing of the composite with the largest \ion{Mg}{2} shift.
Once again, the ``shift'' does not appear to be a shift of the
centroid so much as a suppression of red flux in those emission lines
that appear shifted.  Note that without follow-up observations, we
have no way of knowing if the quasars with the largest \ion{Mg}{2}
blueshifts also have the largest \ion{C}{4} blueshifts (although this
would be a reasonable assumption).

A more thorough analysis of the emission line blueshifts would include
gravitational redshifts \markcite{pva+95}(e.g., {Popovic} {et~al.} 1995).  However, we believe
that they are probably not important, for a number of reasons.  First,
they are relatively small; using values chosen to maximize the
graviational redshift ($M_{\rm BH} = 10^9\,M_{\odot}$ and a distance
of 1 light-month between the black hole and the BELR) yields a
gravitational reshift of only $\sim200\,{\rm km\,s^{-1}}$
\markcite{pva+95}({Popovic} {et~al.} 1995).  Second, unless the blueshift is correlated with the
mass of the central black hole, then there should not be a correlation
between the emission line blueshifts and the strength of a
gravitational redshift.  Finally, and most importantly, they are in
the wrong direction: it is when \ion{C}{4} has the largest {\em
blueshift} that the \ion{C}{4} redshift is least consistent with the
redshift from narrow, forbidden emission features.

\subsection{Correlations with Other Properties}

We now turn to a comparison of the \ion{C}{4} velocity shift
composites as a function of a number of quasar properties.
Tables~\ref{tab:tab1} and~\ref{tab:tab2} give the properties of the
four \ion{C}{4} shifted composite spectra.  Each of the four composite
spectra has a similar number of input spectra.  We find that the four
composites have relatively similar redshift and luminosity
distributions, consistent with there being no strong correlation
between these properties and the velocity shift of the \ion{C}{4}
emission line.  We further tested this conclusion by creating
luminosity and redshift composite spectra and found no significant
emission line shifts between high- and low-luminosity quasars and
between high- and low-redshift quasars.  The redshift range covered by
these composites is probably not large enough to draw any significant
conclusions, however.  On the other hand, the range in absolute
magnitude, $-24.54 \ge M_i \ge -28.57$, is sufficient to test any
significant trend with quasar luminosity.  Although the distributions
appear similar, there is a small trend of increasing \ion{C}{4}
blueshift with luminosity; see \S~\ref{sec:baldwin} for further
discussion.

We {\em do} find that there is a possible trend in the radio
properties as a function of the \ion{C}{4} velocity shift.  Greater
\ion{C}{4} velocity shifts are correlated with lower radio detection
fraction as can be seen in Table~\ref{tab:tab2}; this finding is
consistent with the results of \markcite{msd+96}{Marziani} {et~al.} (1996).  Although the absolute
numbers are small, we find that composite A has a $3\sigma$ excess
(using the statistical definitions of \markcite{geh86}{Gehrels} 1986) of radio
detections as compared to composite D.  In addition, we find that
there is a trend in radio flux density; the radio-detected quasars are
less radio bright with increasing \ion{C}{4} blueshift; the four
sample D objects have an average flux density of only $1.27\,{\rm
mJy}$.  To test that our results are not biased by the inclusion of
more radio sources in sample A as compared to sample D, we have also
re-created the composite spectra after removing all of the radio
detected quasars.  We find that the removal of these spectra produces
no qualitative differences between the original and non-radio
composite spectra.

Unfortunately, there are not enough radio-detected quasars in our
sample (only 41 of the 794 in the \ion{C}{4}/\ion{Mg}{2} sample, many
of which are only faint radio sources) to test if there is any
correlation between \ion{C}{4} velocity shift and quasar orientation
as measured by radio spectral index or radio core dominance.
Nevertheless it is interesting to note that the SDSS+FIRST quasars in
our sample are strongly biased against lobe-dominated quasars, since
the SDSS Quasar Survey \markcite{ric+02}({Richards et al.} 2002) requires that FIRST-selected
quasars have a match to an SDSS optical source within $2\arcsec$.
This restriction will cause the SDSS to fail to target (as
radio-selected quasars) those objects which have strong radio lobes,
but no core radio emission (many of these objects will still be
targeted as color-selected quasar candidates).

Although we would expect there to be unmatched double-lobe radio
sources in our sample, a search of the FIRST database out to
$120\arcsec$ and an examination of the images does not reveal many
cases, and there is no obvious trend with \ion{C}{4} blueshift.
However, it may be that the high resolution of the FIRST survey causes
faint lobe emission to be missed.  Possible evidence in support of a
lack of lobe-dominated sources is the fact that of the 18
radio-detected quasars in our sample that are brighter than $25\,{\rm
mJy}$ at $20\,{\rm cm}$, 12 are clearly core dominated as determined
by visual inspection of the FIRST \markcite{bwh95}({Becker} {et~al.} 1995) images, and only one
of the other six has clear radio lobes.  Thus, we might draw the
conclusion that there is a dearth of radio-detected quasars in sample
D because there is a selection effect that causes a correlation
between large \ion{C}{4} velocity shift and lobe-dominated quasars
(which are thought to be seen edge-on to the plane of the disk).  See
\S~\ref{sec:orientation} for further discussion.

We find a similar correlation with the X-ray detections as determined
from the EDR quasar catalog \markcite{sch+02}({Schneider} {et~al.} 2002); however, it is
attributable to the fact that quasars from composite A, for whatever
reason, have more pointed ROSAT observations.  There is no trend when
considering only the ROSAT Faint Source Catalog sources
\markcite{vab+00}({Voges} {et~al.} 2000).  It would be interesting to examine the X-ray
properties in more detail.  Specifically, if the velocity shift effect
is caused by orientation, and if the correlation between radio
core-dominance and soft X-ray energy index is real \markcite{swe+93}({Shastri} {et~al.} 1993),
then we might expect that the quasars in sample A will have flatter
soft X-ray spectral indices than the quasars in sample D.

\section{Discussion \label{sec:discussion}}

\subsection{BELR Models}

The results presented herein will help to constrain models for the
BELR, in particular, it should be clear that there cannot be any doubt
as to the reality of the systematic blueshift of \ion{C}{4} emission
lines; this shift {\em must} be accounted for in any AGN models.  This
could be considered a strike against the model of \markcite{elv00}{Elvis} (2000), in
which the blueshift of high-ionization emission lines relative to
low-ionization emission lines has ``no obvious interpretation''.
However, it is equally important that these models incorporate the new
results herein, primarily that the blueshift does not seem to be a net
blueshift, but rather a blueshift resulting from a lack of flux in the
red wing of the emission line.

For cloud models, the similar blue wings of \ion{C}{4} emission in the
different composites would appear to favor an intrinsically isotropic
velocity field in an isotropically-emitting BELR\footnote{It has been
argued that the lack of short time delay response in reverberation
mapping studies appears to rule out spherically symmetric models, but
this is not the case if the emission from the BEL clouds is
anisotropic \markcite{fph+92}({Ferland} {et~al.} 1992), which is very likely.}, plus obscuration
that preferentially removes flux from the red wing of \ion{C}{4}.
However, such preferential obscuration may be difficult to produce.
Unfortunately, the parameter space available to BELR models is huge,
and it becomes difficult to distinguish between using observations to
constrain a model and fine tuning a model to match the observations ex
post facto.

Thus, a detailed discussion of how our results constrain BELR models
is beyond the scope of this paper.  However, we briefly comment on the
subject.  First, we make two general assumptions: 1) that the BELR
region is stratified in ionization, specifically that the
low-ionization region is located farther from the continuum source
than the high-ionization region \markcite{pet93}({Peterson} 1993), but see \markcite{odl+98}{O'Brien} {et~al.} (1998);
and 2) that an optically thick accretion disk is interior to both
regions.  We then hypothesize that the differences between composite A
and D are an orientation effect with composite A representing a
face-on configuration and composite D representing a more edge-on
configuration.

\begin{figure}[p]
\epsscale{1.0} 
\plotone{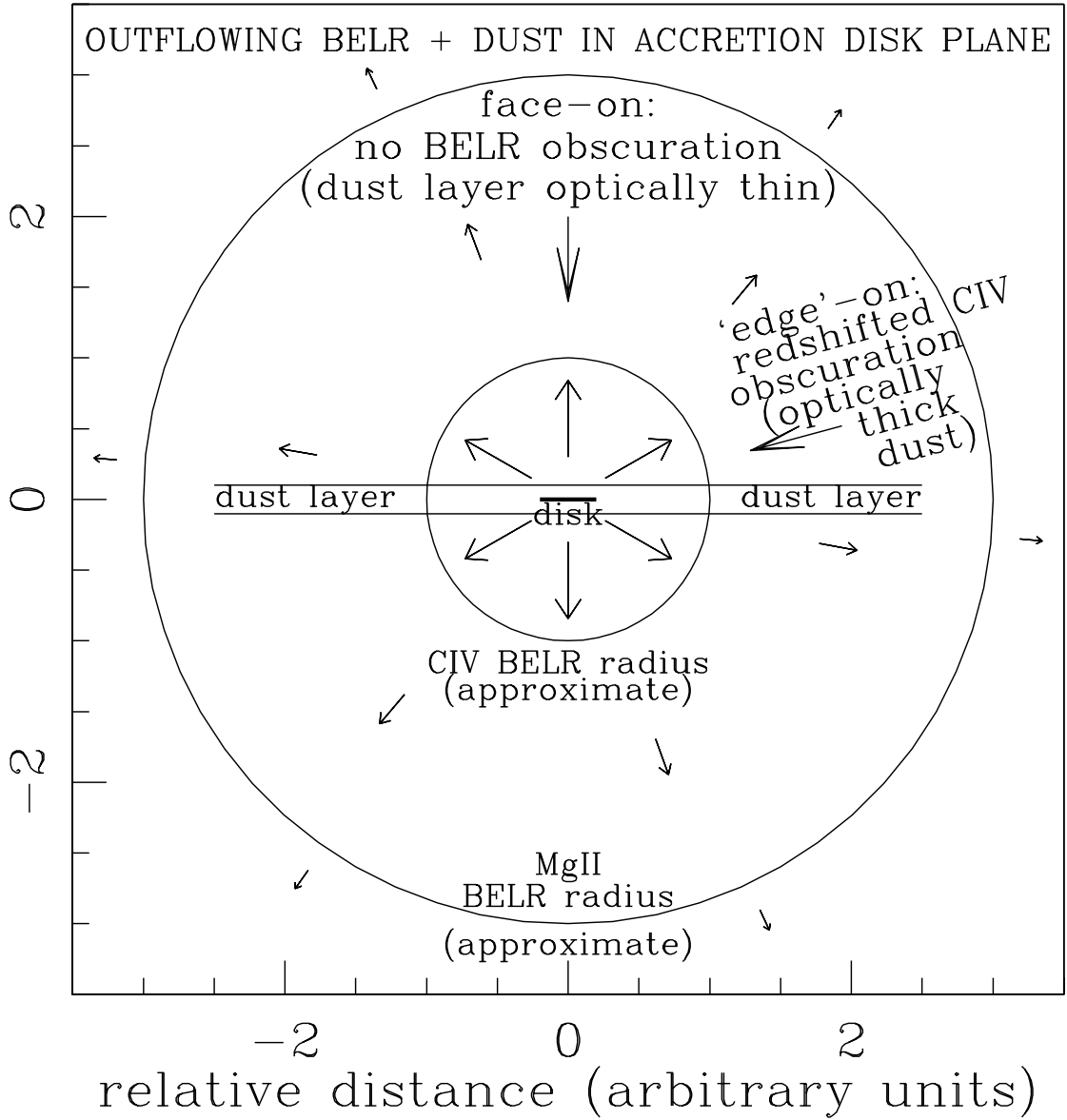}
\caption{Schematic diagram of the BELR in an outflowing cloud
scenario.  The units are arbitrary, but the size of the \ion{C}{4}
BELR relative to the \ion{Mg}{2} BELR is scaled by about the same
factor as the relative reverberation mapping delay times for each line
\markcite{pet93}({Peterson} 1993). \label{fig:fig9}}
\end{figure}

Further assuming a cloud model for the BELR with a roughly isotropic
distribution of {\em outflowing} clouds, one possible (but perhaps
unlikely) scenario could involve an equatorial screen that is
optically thin in the face-on direction, but becomes optically thick
to clouds on the side of the accretion disk facing away from the
observer in the edge-on direction (Figure~\ref{fig:fig9}).  The
high-ionization clouds are assumed to be outflowing faster than the
low-ionization clouds, and the dust layer is assumed to be located
mostly interior to the typical \ion{Mg}{2} emission radius.  Such a
``model'' could produce emission profiles similar to those that we
observe.  However, if the clouds were infalling, it is difficult to
imagine a way for obscuration to produce the observed results.

We must also consider disk-like BELR geometries, such as discussed by
\markcite{kk94}{Konigl} \& {Kartje} (1994), \markcite{bks+97}{Bottorff} {et~al.} (1997), \markcite{mc98}{Murray} \& {Chiang} (1998), and others.  If we invoke
a disk-wind type model instead of a spherically symmetric cloud model,
then the situation may be quite different.  In this case, the flow is
not purely radial: the velocity differences can be due purely to
orientation and a screen is not needed.  However, since we observe
that the emission line profiles are not simply shifted, but also they
are weaker, a screen may still be necessary.  For such a disk-wind
model the most obvious source of a blueshift would be an outflow in
the plane of the disk.  In this case, a face-on orientation would
yield no net shift, whereas an edge-on orientation could produce a net
shift if the far side of the disk were obscured.  However, we would
still have to account for the fact that our observations suggest that
there is not a net blueshift, just an apparent blueshift.  In
addition, although our hypothesis assumes that the proposed
orientation effect is a result of the orientation of the disk itself,
it could instead result from a change in the opening angle of the
disk-wind, see \S~\ref{sec:orientation} for further discussion.

We note that \markcite{mc97}{Murray} \& {Chiang} (1997) were able to use a varying outer radius for
the emission line gas in a disk-wind model to explain the emission
profiles in \markcite{wbf+93}{Wills} {et~al.} (1993). \markcite{mc97}{Murray} \& {Chiang} (1997) also found that additionally
varying the inclination angle did not produce a better fit.  However,
the \markcite{wbf+93}{Wills} {et~al.} (1993) composites all place the peak of the \ion{C}{4}
emission profile at zero velocity.  Since the \ion{C}{4} emission line
profiles of our four composites are so similar to theirs, it seems
quite likely that their ``composite 4'' should be blueshifted in a
similar manner to our ``composite D'' (and so on for the other
composites).  Thus, a varying outer radius cannot be the whole
picture, and it would be interesting to re-address this issue.

In any case, it should be obvious that these observations and further
study of both cloud and disk-wind models, combined with the results of
reverberation mapping data and analysis \markcite{kab+95,dk96}(e.g., {Korista} {et~al.} 1995; {Done} \& {Krolik} 1996)
will lead to a significantly better understanding of the broad line
region in AGN.

\subsection{The Baldwin Effect \label{sec:baldwin}}

The Baldwin Effect was originally defined as an anti-correlation
between the equivalent width of \ion{C}{4} emission and the continuum
luminosity at $1450\,{\rm \AA}$ in the spectra of quasars
\markcite{bal77}({Baldwin} 1977).  While the original definition is still the dominant
one, the Baldwin Effect has come to mean any anti-correlation between
quasar luminosity or absolute magnitude and the strength of the broad
emission lines.  Since our sample shows a very clear evolution of
\ion{C}{4} equivalent width from large equivalent width in sample A to
a much smaller equivalent width in sample D, a discussion of how the
Baldwin Effect might relate to our results is warranted.  Although the
mean absolute magnitudes of our sample and a histogram of the
individual absolute magnitudes in each sample (not shown) are quite
similar, we were surprised to find that there is a significant
difference between the samples in terms of their absolute magnitudes.
The mean values of $M_{i^*}$ for samples A and D are $-26.14$ and
$-26.46$, respectively.  This difference has a Student's $t$
\markcite{ptv+92}({Press} {et~al.} 1992) of $4.46$ with a probability of $1.05\times10^{-5}$,
indicating a significant difference between the samples.  The
differences between the adjacent samples (e.g., A and B) are only
marginally significant.

To determine if our results are influenced by luminosity, we have
created control samples by constraining the absolute magnitude
distributions for sub-samples of each sample.  We first bin the
absolute magnitudes in bins of 0.5 mag, then count how many quasars
are in each bin for each sample.  We then created a sub-sample for
each sample with only the minimum number of quasars in each bin.  In
this way, we create sub-samples with the same absolute magnitude
distributions.  A Student's $t$ test confirms the similarity of the
sub-samples in terms of their mean absolute $i$-band magnitudes.
Using these absolute magnitude normalized subsamples, we have
recreated Figure~\ref{fig:fig4} and find no qualitative differences as
compared to the full samples.  We thus conclude that changes in
$M_{i^*}$ are not driving the differences between the original
samples.

Nevertheless, a better analysis in terms of the Baldwin Effect is
still needed.  Our $M_{i^*}$ values sample both the continuum and
emission line flux at $\lambda\sim2800\,{\rm \AA}$, whereas the
Baldwin Effect is usually in terms of the continuum flux at
$\lambda\sim1450\,{\rm \AA}$.  More work is clearly needed to
understand the exact relationship between our observations and the
Balwin Effect.

It is particularly interesting to note that the lines that show the
most differences between samples A and D are also the ones that have
the strongest reported Baldwin Effects, in particular, \ion{C}{4} and
\ion{He}{2} \markcite{opg94}({Osmer}, {Porter}, \& {Green} 1994).  We also see no strong differences in the
\ion{Si}{4} profiles or blueshifts; no Baldwin Effect is observed for
this line \markcite{opg94}({Osmer} {et~al.} 1994).  These observations are unusual in the sense
that both \ion{C}{4} and \ion{Si}{4} are high ionization lines with
very similar chemical structure.

Finally, if the Baldwin Effect and the blueshifts of quasar emission
lines are related, one must take the blueshifts into account when
determining the luminosity of the quasars.  Large blueshifts will
cause the redshift and thus the luminosity of the quasar to be
systematically underestimated.  Since we have shown that those quasars
with large blueshifts also have weaker \ion{C}{4} emission, large
emission line blueshifts will cause the the Baldwin Effect to appear
weaker than it should.

\subsection{Broad Absorption Line Quasars}

Since the \ion{C}{4} troughs in BAL quasars removes some of the
\ion{C}{4} emission flux, it is nearly impossible to ask whether
individual BAL quasars have small or large \ion{C}{4} emission line
blueshifts.  However, we can study how the ensemble average of BALs
compares to the composites that we have created.  First, we divide the
BALs into two types: low-ionization (LoBALs) and high-ionization
(HiBALs), where, in addition to absorption by \ion{C}{4}, LoBALs show
absorption in \ion{Mg}{2} and other low-ionization lines.  The HiBAL
composite was created by combining all of the non-LoBAL BAL-like
objects that were originally rejected from our \ion{C}{4} sample; the
redshifts are all taken from \ion{Mg}{2} as they were for our
\ion{C}{4} sample.  The LoBAL composite includes the LoBALs from the
HiBAL sample and includes some LoBALs outside of the redshift range
studied in the \ion{C}{4} -- \ion{Mg}{2} sample in order to improve
the signal-to-noise of the composite spectrum.  These BAL composite
spectra were created in exactly the same manner as composites A, B, C,
and D.

The composite BAL spectra were then scaled as follows.  A composite
spectrum \markcite{vrb+01}({Vanden Berk} {et~al.} 2001) was fitted to each of them in the rest
wavelength frame by minimizing a weighted $\chi^2$ merit function,
normalized in the region $1725\pm25\,{\rm \AA}$.  The spectral index
and reddening, ${\rm E}(B-V)$, were allowed to vary, using a reddening
law with
\begin{equation}
f_\lambda \propto \lambda^{\alpha_\lambda} e^{-2{\rm E}(B-V)/\lambda},
\end{equation}
(see \markcite{fww00}{Francis}, {Whiting}, \& {Webster} 2000), where $\lambda$ is in $\mu{\rm m}$ and
$\alpha_\lambda = -(2+\alpha)$.  The $\chi^2$ function was weighted at
certain wavelengths to eliminate or reduce the effects of several
emission and absorption features.  The weight was 0 for the major
emission lines from Lyman-$\alpha$ to \ion{Mg}{2}, and also for the
Lyman-$\alpha$ forest; 0.5 for the region between \ion{C}{4} and
\ion{Si}{4} where BAL troughs are most likely to be found; and 1 for
other wavelengths.  The chi-square function was then minimized using a
quasi-Newton method, yielding best-fit values of the spectral index
and reddening.  See Reichard et al. (2002, in preparation) for a more
detailed discussion.

\begin{figure}[p]
\epsscale{1.0} 
\plotone{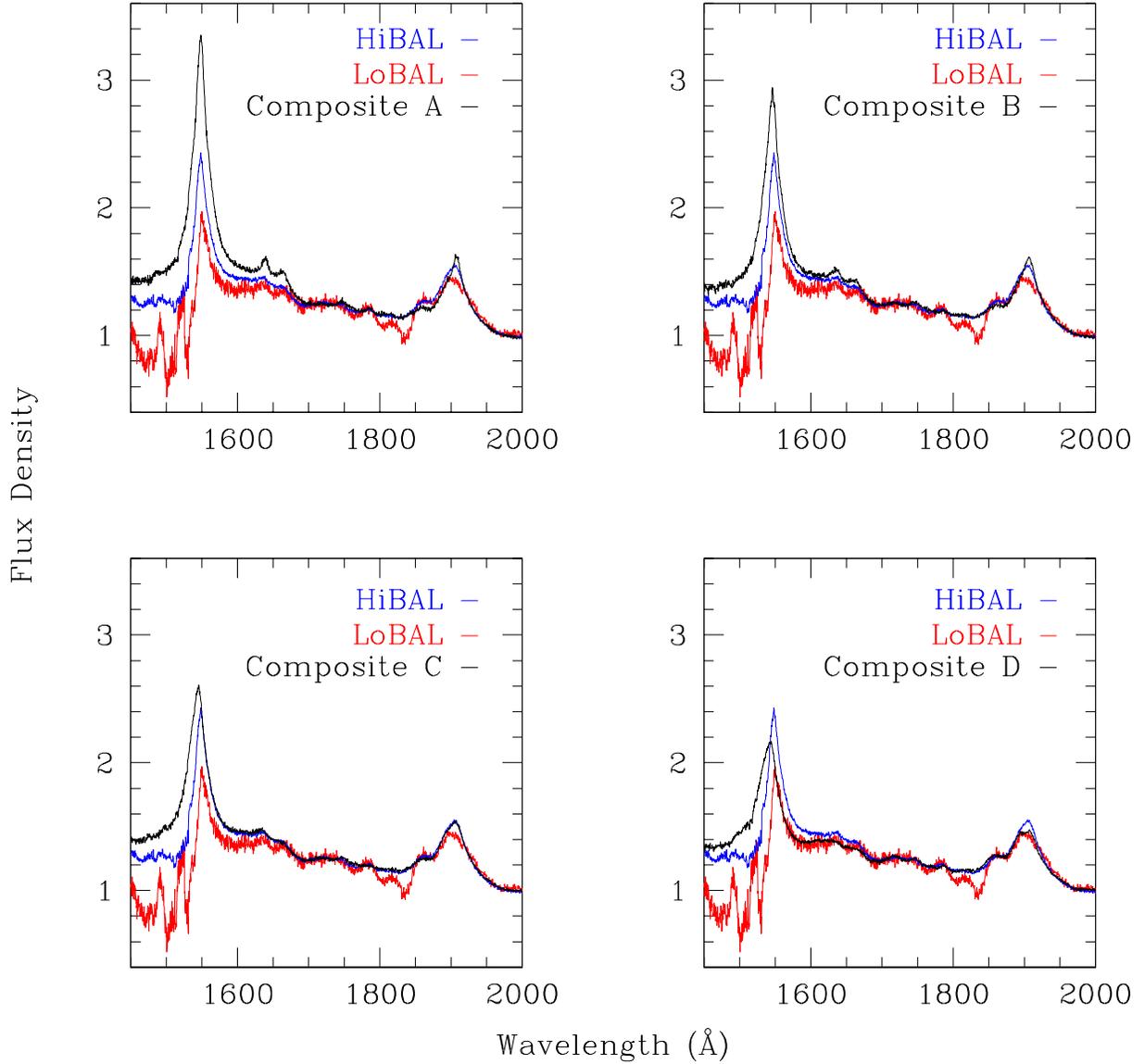}
\caption{High-ionization BAL composite spectrum (HiBAL; blue line) and
low-ionization BAL composite spectrum (LoBAL; red line) overlayed on
each of our four composites.  The \ion{C}{4} and \ion{C}{3}] emission
lines in the HiBAL composite most closely resemble composite C,
whereas the \ion{C}{4} and \ion{C}{3}] profiles in the LoBAL composite
most closely resemble those in composite D. \label{fig:fig10}}
\end{figure}

In order to compare the BAL composites to our A, B, C, and D
composites, we also scale composites A, B, C, and D in the same manner
as above.  Thus the composites from samples A, B, C, and D will have
roughly the same (overall) continuum slope and shape as the BAL
composites, which allows comparison of their emission line features.
From Figure~\ref{fig:fig10}, it is quite clear that the LoBAL
composite is most similar to composite D: the red wing of \ion{C}{4}
matches quite well as does the \ion{C}{3}] blend.  In addition, the
LoBAL composite also shows a clear weakening of \ion{He}{2} emission.
The HiBAL composite is most similar to composite C; the \ion{C}{4} and
\ion{C}{3}] emission line profiles are good matches.  Also note that,
based upon the upper left-hand panel, it would appear that the
\ion{C}{4} BAL outflows begin right at zero velocity with respect to
the systemic redshift (as defined by \ion{Mg}{2}).

Since the viewing angles of BALs are generally thought to be
constrained to a relatively small angle from the accretion disk plane,
that the LoBAL composite best matches composite D is consistent with
the idea that composite D is composed of quasars that are
preferentially observed in an edge-on configuration.  The fact that
the HiBAL composite is most similar to composite C is also consistent
with the orientation hypothesis: HiBALs may differ from LoBALs in that
the line of sight passes through somewhat less dense gas clouds that
are more highly ionized.  These correlations would seem to be
inconsistent with, but do not necessarily rule out, the idea that BALs
may (also) be an evolutionary stage of all quasars \markcite{bwg+00}({Becker} {et~al.} 2000).

\subsection{Associated Absorption}

Not only does the blueshift of the \ion{C}{4} emission line affect the
study of BAL absorption, it also affects our understanding of narrow,
``associated'' absorption \markcite{fwp+86}({Foltz} {et~al.} 1986).  These are absorption
systems that are preferentially within $\pm3000\,{\rm km\,s^{-1}}$ of
the quasar redshift; they occur at a much higher rate (per unit
redshift) than absorbers which are thought to be caused by intervening
galaxies at redshifts much smaller than the quasars.  Since their
discovery, two hypotheses have dominated the possible explanations,
\markcite{fwp+86}{Foltz} {et~al.} (1986) suggested that 1) they are due to outflows, similar to,
but weaker than BALs, or 2) they are due to a virialized cluster-like
velocity dispersion.  However, the fact that the \ion{C}{4} emission
line can be blueshifted changes the situation significantly
\markcite{mrr+99}({McIntosh} {et~al.} 1999).

\begin{figure}[p]
\epsscale{1.0}
\plotone{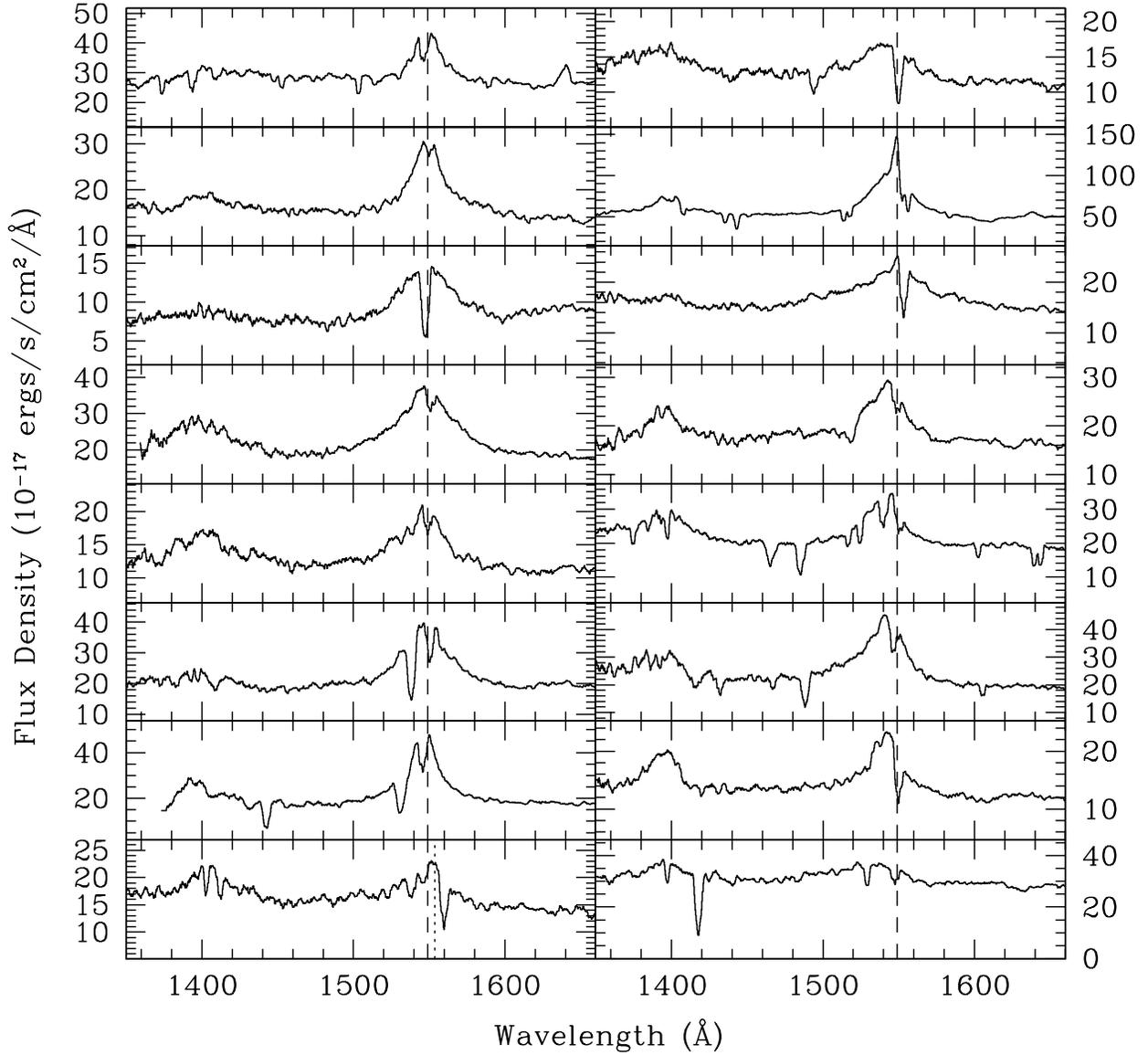}
\caption{\ion{C}{4} emission line region of quasars with maximally
redshifted (``infalling''), very strong associated absorption.  Dashed
line shows the expected peak of \ion{C}{4} based upon the \ion{Mg}{2}
redshift.  ({\em Left}) Quasars from sample A, with small \ion{C}{4}
emission line blueshifts.  Dotted line in the lower left-hand panel
shows the expected location of \ion{C}{4} based on the hand-measured
\ion{Mg}{2} redshift for this quasar.  ({\em Right}) Quasars from
sample D, with large \ion{C}{4} emission line
blueshifts.\label{fig:fig11}}
\end{figure}

A full analysis of this issue is beyond the scope of this work;
however, we have performed a simple test that helps to shed light on
this subject.  One of us (GTR) examined the \ion{C}{4} emission line
regions of all the spectra that comprise composites A and D, and two
samples of quasars (one each from samples A and D) with good
signal-to-noise and strong associated absorption were constructed.
These samples were further restricted to those objects showing the
largest associated absorption redshifts (i.e., towards or in the red
wing of \ion{C}{4}).  We present the \ion{C}{4} emission line regions
of some of the quasars from these biased sub-samples in
Figure~\ref{fig:fig11}.  The quasars on the left-hand-side are from
sample A and have small \ion{C}{4} emission line blueshifts with
respect to \ion{Mg}{2}, whereas the quasars on the right-hand-side are
from sample D and have larger \ion{C}{4} emission line blueshifts with
respect to \ion{Mg}{2}.

What is interesting about Figure~\ref{fig:fig11} is that, in almost
every case, the associated absorption is at or shortward of the
systemic redshift (dashed line, as measured from \ion{Mg}{2}), and
thus are consistent with outflows.  Large virial velocities are not
necessary to explain the ``infalling'' absorbers.  Furthermore, in
only one sample A case does the absorption occur entirely in the red
wing of the emission profile.  For this one exception, we suspect that
a \ion{C}{4} blueshift that has a larger error than normal is to
blame; when we remeasured the \ion{Mg}{2} redshift by hand, we found
that the redshift from the automated code was too small by
$\sim900\,{\rm km\,s^{-1}}$.  The expected location of \ion{C}{4}
using the revised redshift is indicated by the dotted line in the
bottom panel on the left-hand side of Figure~\ref{fig:fig11}.
However, given the expected errors in the individual measurements, it
it unlikely that the errors will produce a symmetric distribution
around the \ion{C}{4} line center, but quite possible that the errors
could explain any of the remaining apparent inflows.  That is, if the
velocity shifts of associated absorption lines were always referenced
to the systemic redshift, we might find that associated absorbers are
all outflows.

Since \markcite{fwp+86}{Foltz} {et~al.} (1986) and others \markcite{rlb+01}(e.g., {Richards} {et~al.} 2001) have found
that the strong associated absorbers are predominantly found in
steep-spectrum radio-loud quasars, it is interesting to think about
them in terms of an orientation hypothesis.  First, we note that the
presence of associated absorption in our sample A quasars may be an
argument {\em against} an orientation interpretation for the
\ion{C}{4} emission line blueshifts.  Steep-spectrum quasars are
thought to have a more edge-on orientation, whereas flat-spectrum
quasars are thought to be more face-on.  Thus the observation that
associated absorbers are predominantly found in steep-spectrum quasars
and that our hypothesized face-on quasars can have strong associated
absorption is apparently inconsistent.

However, it may be possible to reconcile this problem in the following
manner.  If we define ``associated absorption'' as any absorption that
occurs within the profile of \ion{C}{4} (say within $\pm3000\,{\rm
km\,s^{-1}}$ of the emission line peak --- regardless of whether the
peak is blueshifted or not), {\em and} we assume that the associated
absorbers are purely outflows, then we would expect to see a higher
density of associated absorption per unit velocity in the objects in
sample D (assuming that they are seen edge-on).  This is because there
is a bigger range of velocities in which the absorbers can be found in
sample D quasars, since the \ion{C}{4} emission line is blueshifted.
For example, in a quasar with no blueshift of the \ion{C}{4} emission
line, the observable range of associated absorption velocities is only
$3000\,{\rm km\,s^{-1}}$, but in a quasar where the \ion{C}{4}
emission line is blueshifted by $3000\,{\rm km\,s^{-1}}$, the
observable range of associated absorption is $6000\,{\rm km\,s^{-1}}$;
in both cases we {\em assume} that the full range over which
associated absorption can be seen is $6000\,{\rm km\,s^{-1}}$, so a
sample of quasars with blueshifted emission lines would be expected to
have a higher density of associated absorption.  If sample D objects
tend to be steep-spectrum, lobe-dominated objects when they are radio
sources, this could explain the excess of associated absorbers seen in
steep-spectrum, lobe-dominated radio-loud quasars.

If associated absorbers are really outflows, then they may be present
at velocities much larger than $3000\,{\rm km\,s^{-1}}$ from the
emission redshift of the quasar, as was suggested by \markcite{rlb+01}{Richards} {et~al.} (2001).
Thus, there may be a large population of narrow, intrinsic absorption
systems in flat-spectrum face-on quasars, see \markcite{rlb+01}{Richards} {et~al.} (2001) and
references therein.

\subsection{An Orientation Effect? \label{sec:orientation}}

What causes the blueshifted emission lines in quasars and other
similar phenomena (e.g., the Baldwin Effect)?  Orientation,
specifically orientation of the plane of the accretion disk is
certainly one possibility.  We have presented some results that are
consistent with this hypothesis.

We have shown that BAL quasars are most similar to the quasars from
samples C and D.  Since most models of BALs have the BAL clouds
confined to a region within a small angle of the disk (or the
disk-wind), this observation may be consistent with an orientation
effect.  The case for associated absorption, which is preferentially
seen in steep-spectrum (edge-on) quasars, is less clear, but could
also be consistent with an orientation effect.

The radio properties of our sample may also support the hypothesis
that orientation plays a key role in the blueshifting of the
\ion{C}{4} emission line.  We find that there are significantly more
radio-detected quasars in sample A than sample D and that the sample A
objects are brighter at radio wavelengths.  If the SDSS+FIRST matching
algorithm is failing to find lobe dominated quasars, or if FIRST is
failing to detect some lobe dominated quasars, an orientation effect
might explain the lack of radio-detected sources in sample D.
Furthermore that the sample A sources are bright, whereas the sample D
sources are faint could be the result of an orientation-related
beaming effect.

An edge-on orientation for the largest \ion{C}{4} shifts may also be
consistent with the reduced \ion{C}{4} emission and the lack of
\ion{He}{2} emission.  These are both high-ionization lines that are
likely to be formed closer to the source of ionizing radiation than
low-ionization lines and thus would be easier to obscure than the
low-ionization lines which are presumably formed further out.  If the
obscuration is restricted to the plane of the presumed accretion disk,
then we might expect that some of the high-ionization lines would be
completely or partially obscured in an edge-on orientation.

Furthermore, if the disk-wind model for the BELR is true, then we
might expect a broadening of the lines with increasing disk
inclination angle as seen in H$\beta$ by \markcite{wb86}{Wills} \& {Browne} (1986).  \markcite{vwb00}{Vestergaard}, {Wilkes}, \&  {Barthel} (2000)
find that systems which appear to be inclined according to their radio
properties have broader emission line wings, which is consistent with
the material being seen in a preferred direction.  We do indeed see an
increase in the FWHM of \ion{C}{4} with an increase in our proposed
viewing angle (see Table~2).  However, as we discussed earlier, it is
not clear that this broadening of the line is real; it may instead be
due to partial obscuration of the line.

A possible argument {\em against} the orientation hypothesis is the
fact that \markcite{bh95}{Baker} \& {Hunstead} (1995) found significant differences in the strength
of the $3000\,{\rm \AA}$ bump between composite spectra that are core-
and lobe-dominated (core-dominated having stronger $3000\,{\rm \AA}$
bump emission).  In their sample, the $3000\,{\rm \AA}$ bump is barely
visible for steep-spectrum quasars and virtually disappears for {\em
compact} steep-spectrum quasars.  Since core-dominance is thought to
be correlated with orientation in radio-loud quasars, this change in
$3000\,{\rm \AA}$ bump strength may also be related to orientation.
Although we do see some differences in the strength of $3000\,{\rm
\AA}$ bump between composites A and D, they do not appear to be
anywhere near as apparent as those found by \markcite{bh95}{Baker} \& {Hunstead} (1995).  If both our
effects and the \markcite{bh95}{Baker} \& {Hunstead} (1995) effects are caused by orientation, we
might expect to see a bigger difference in the \ion{Fe}{2} fluxes
between composites A and D.

Although the unified model \markcite{up95}(e.g., {Urry} \& {Padovani} 1995) is well-established,
more and more emphasis is being placed on properties other than
orientation in terms of explaining differences between quasars.  These
include luminosity, the Eddington ratio, accretion rate, black hole
mass, black hole spin and age \markcite{lao00}(e.g., {Laor} 2000).  For example,
\markcite{bg92}{Boroson} \& {Green} (1992) argue that the first eigenvector from their principal
component analysis (``Eigenvector 1'' --- which illuminates some
correlations between intrinsic quasar properties) is driven by
something other than external orientation.  Even though the properties
of our sample are very similar to the ``Eigenvector 1'' properties
(see \markcite{bf99}{Brotherton} \& {Francis} 1999, Table 2), we prefer orientation as an
explanation.  However, this orientation effect need not be specific to
the orientation of the accretion disk (i.e. external orientation), but
rather it could be related to the opening angle of a disk wind (such
as in \markcite{elv00}{Elvis} 2000, Figure 7), where some other quasar property is
causing the opening angle of the disk wind to change, thus producing
an orientation type effect (i.e. internal orientation).  A strong
argument for this type of behavior (or against orientation entirely
depending on how you look at it) is the intrinsic Baldwin Effect
\markcite{pp92}({Pogge} \& {Peterson} 1992).  Since a Baldwin type effect is seen in individual
quasars as their continuum luminosity varies (but with a different
slope than the ``external'' Baldwin Effect), the Baldwin Effect cannot
be entirely due to the orientation of the accretion disk; the same is
true for the emission line blueshifts if they are related to the
Baldwin Effect.  However, if the orientation effect is really the wind
opening angle or the angle the wind makes with the disk axis and these
angles can change on reasonably short timescales, then this might
account for the major differences in quasar spectra.

In any case, it is certainly possible that the size of the \ion{C}{4}
emission line blueshift may be a function of the orientation of the
quasar.  Whether the orientation in question is external or internal
remains to be seen.  Further study, particularly with regard to radio
and X-ray properties, is needed to determine if this is really the
case.

\subsection{Future Work}

Some areas of future work are beyond the scope of this paper, but are
clearly appropriate, including the following:

1) Ideally, we would like to know the velocity offset between the
   \ion{C}{4} and [\ion{O}{3}] emission lines, which requires that we
   obtain IR spectra of the [\ion{O}{3}] emission line region for
   those quasars where we already cover \ion{C}{4} and \ion{Mg}{2}.

2) Similarly, we should attempt to extend our low redshift sample to
   cover the \ion{C}{4} emission line region by obtaining UV spectra
   of these quasars.

3) UV spectra would also be appropriate for those objects where we
   cannot see the Lyman-$\alpha$ emission line, since the profile of
   Lyman-$\alpha$ may be helpful in constraining BELR models
   \markcite{kk86}({Kallman} \& {Krolik} 1986).

4) If we could find a way to know how large the \ion{C}{4} blueshift
   is likely to be without looking at the \ion{C}{4} emission line, it
   would be possible to study how the \ion{C}{4} blueshifts correlate
   with broad absorption line properties.

5) A detailed study of associated absorbers among the quasars studied
   herein is clearly in order.  In many cases, this will mean getting
   higher signal-to-noise spectra.

6) Since the \ion{C}{4} blueshifts so strongly correlate with the
   equivalent width of the emission lines, a more detailed
   investigation of how this affects the Baldwin Effect is in order.

7) It is possible that emission from \ion{Fe}{2} and \ion{Fe}{3} could
   be influencing some of the changes that we attribute to other
   species.  Removing the iron emission complexes using the
   \markcite{vw01}{Vestergaard} \& {Wilkes} (2001) iron template would address this issue.  We have not
   chosen to attempt this procedure here, in part because the
   \markcite{vw01}{Vestergaard} \& {Wilkes} (2001) iron template is not complete in the critical regions
   near \ion{C}{4} and \ion{Mg}{2} emission.

8) Last, but perhaps most important, is that future AGN ``models''
   must not fail to take into account these emission line blueshifts.
   Furthermore, they must be able to explain the emission line
   profiles.  These should be key ingredients to any model.

\section{Conclusions}

Using two samples of quasars from the SDSS Early Data Release Quasar
Catalog \markcite{sch+02}({Schneider} {et~al.} 2002), we have studied the differences in redshift of
quasars as measured by different emission lines.  In one sample (with
794 quasars in the redshift range, $1.54 \le z \le 2.2$), we compared
the redshifts as determined from the peak of the \ion{C}{4} emission
line to those as determined from the peak of the \ion{Mg}{2} emission
line.  In the second sample (with 417 quasars in the redshift range,
$0.415 \le z \le 0.827$), we compared the redshifts from \ion{Mg}{2}
to the redshifts from [\ion{O}{3}].  Our conclusions based on these
two samples can be succinctly summarizes as follows:

1) The \ion{C}{4} emission line is not typically found at the same
   redshift as \ion{Mg}{2}, a fact which has been reported previously.
   The shift of \ion{C}{4} is typically $\sim800\,{\rm km\,s^{-1}}$,
   but can be as large as $3000\,{\rm km\,s^{-1}}$.

2) This shift is apparently not a bulk shift, but rather is due to a
   lack of flux in the red wing of the \ion{C}{4} emission line
   profile.  This fact is a new result, but is not entirely without
   historical precedent.  Although this conclusion depends on the
   placement of the continuum, we feel that our handling of the
   continuum is appropriate and that this conclusion is robust.

3) The equivalent width of the \ion{C}{4} emission line decreases with
   increasing blueshift.

4) The FWHM of the \ion{C}{4} emission line increases with increasing
   blueshift of the peak of the emission line; however, this effect
   may not be real.  For example, if it is true that the red wing is
   simply attenuated for sample D quasars, then the FWHM will not have
   the same meaning that it does if there is no attenuation.

5) The blueshift of the \ion{C}{4} emission line may correlate with
   the orientation of the quasar as measured by radio properties
   (large blueshift having an equatorial viewing angle); further work
   is clearly needed.

6) The apparent blueshift of the \ion{C}{3}] emission line with
   increasing blueshift of the \ion{C}{4} emission line may instead be
   the result of varying flux ratios of the blended \ion{C}{3}] and
   \ion{Si}{3}] emission lines.

7) The \ion{Mg}{2} emission line has a small blueshift with respect to
   [\ion{O}{3}] in the ensemble average, but individual deviations can
   be as large as $\pm 500\,{\rm km\,s^{-1}}$.

8) The spectra of BAL quasars appear to be most similar to the spectra
   of quasars with large \ion{C}{4} emission line blueshifts.  If the
   \ion{C}{4} emission line blueshift is correlated with orientation,
   this supports the conclusion that BALs are correlated with
   orientation (whether external or internal), and vice versa.

9) It is possible that redshifted ``associated'' absorption population
   may be explained by the blueshift of \ion{C}{4} emission (as has
   been suggested before).  Even if there are inflows, we find no need
   for extremely large velocities in the ``virialized'' cluster
   material explanation.

\acknowledgements

We acknowledge helpful discussions with Mike Eracleous, Julian Krolik,
Arieh K\"{o}nigl, Nahum Arav, Niel Brandt, Marianne Vestergaard, and
Brad Peterson.  We also thank Bev Wills, Michael Brotherton, and Sarah
Gallagher for their comments on the manuscript.  We thank the
anonymous referee for suggestions that helped improve the paper,
particularly for the suggestion that we include more discussion of the
Baldwin Effect.  GTR, TAR, and DPS acknowledge support from NSF grant
AST99-00703.  PBH acknowledges financial support from Chilean grant
FONDECYT/1010981 and a Fundaci\'{o}n Andes grant.

Funding for the creation and distribution of the SDSS Archive has been
provided by the Alfred P. Sloan Foundation, the Participating
Institutions, the National Aeronautics and Space Administration, the
National Science Foundation, the U.S. Department of Energy, the
Japanese Monbukagakusho, and the Max Planck Society. The SDSS Web site
is http://www.sdss.org/.  The Participating Institutions in the SDSS
are The University of Chicago, Fermilab, the Institute for Advanced
Study, the Japan Participation Group, The Johns Hopkins University,
the Max-Planck-Institute for Astronomy (MPIA), the
Max-Planck-Institute for Astrophysics (MPA), New Mexico State
University, Princeton University, the United States Naval Observatory,
and the University of Washington.

\clearpage



\clearpage

\clearpage

\clearpage

\clearpage

\clearpage

\clearpage


\begin{deluxetable}{crrrrr}
\tabletypesize{\small}
\tablewidth{0pt}
\tablecaption{\ion{C}{4} Shift Composite Spectra Velocity Shift Data \label{tab:tab1}}
\tablehead{
\colhead{Sample} &
\colhead{${\rm N_{QSOs}\tablenotemark{a}}$} &
\colhead{$\overline{v}_{\rm comp.}\tablenotemark{b}$} &
\colhead{$\overline{v}_{\rm ind.}\tablenotemark{c}$} &
\colhead{${v_{\rm min}\tablenotemark{d}}$} &
\colhead{${v_{\rm max}\tablenotemark{e}}$} \\
\colhead{} &
\colhead{} &
\colhead{$({\rm km\,s^{-1}})$} &
\colhead{$({\rm km\,s^{-1}})$} &
\colhead{$({\rm km\,s^{-1}})$} &
\colhead{$({\rm km\,s^{-1}})$}
}
\startdata
A & 199 (160) & 197 & 193 & $-407$ & 456 \\
B & 198 (165) & 606 & 651 & 461 & 824 \\
C & 199 (166) & 1003 & 1039 & 827 & 1261 \\
D & 198 (179) & 1526 & 1596 & 1268 & 1993\tablenotemark{f} \\
\enddata

\tablenotetext{a}{Number of quasars in sample.  Number in parentheses is the number whose positions are within the FIRST survey area \markcite{bwh95}({Becker} {et~al.} 1995).}
\tablenotetext{b}{Average velocity offset of \ion{C}{4} with respect to \ion{Mg}{2} in the composite spectrum.}
\tablenotetext{c}{Average velocity offset of \ion{C}{4} with respect to \ion{Mg}{2} in the individual spectra that constitute the sample.}
\tablenotetext{d}{Minimum velocity offset of \ion{C}{4} with respect to \ion{Mg}{2} in the individual spectra that constitute the sample.}
\tablenotetext{e}{Maximum velocity offset of \ion{C}{4} with respect to \ion{Mg}{2} in the individual spectra that constitute the sample.}
\tablenotetext{f}{Artificial upper limit imposed by the automated analysis code; the true upper limit is $\sim3000\,{\rm km s^{-1}}$.}
\end{deluxetable}

\begin{deluxetable}{crrrcccc}
\tabletypesize{\small}
\tablewidth{0pt}
\tablecaption{\ion{C}{4} Shift Composite Spectra Properties \label{tab:tab2}}
\tablehead{
\colhead{Sample} &
\colhead{$\overline{z}$} &
\colhead{$\overline{M}_{i^*}$} &
\colhead{$\alpha_{\nu}$} &
\colhead{${\rm Radio\;N/\%}$\tablenotemark{a}} &
\colhead{${\rm X-ray\;N/\%}$\tablenotemark{b}} &
\colhead{\ion{C}{4} EQW} &
\colhead{\ion{C}{4} FWHM}
}
\startdata
A & 1.80 & $-26.14$ & $-0.392$ & 21/13.1 & 4/2.0 & 30.27 & 28.17 \\
B & 1.80 & $-26.27$ & $-0.332$ & 11/6.7 & 8/4.0 & 25.75 & 28.51 \\
C & 1.84 & $-26.31$ & $-0.295$ & 5/3.0 & 3/1.5 & 22.43 & 30.43 \\
D & 1.82 & $-26.46$ & $-0.297$ & 4/2.2 & 2/2.0 & 17.96 & 32.40 \\
\enddata
\tablenotetext{a}{Number and fraction of radio detections from the VLA's ``FIRST'' survey \markcite{bwh95}({Becker} {et~al.} 1995).}
\tablenotetext{b}{Number and fraction of X-ray detections includes only the ROSAT Faint Source Catalog detections \markcite{vab+00}({Voges} {et~al.} 2000).}
\end{deluxetable}

\end{document}